\AtBeginDocument{%
  \providecommand\BibTeX{{%
    \normalfont B\kern-0.5em{\scshape i\kern-0.25em b}\kern-0.8em\TeX}}}

\documentclass[sigconf]{acmart}

\copyrightyear{2022}
\acmYear{2022}
\setcopyright{acmlicensed}\acmConference[WWW '22]{Proceedings of the ACM
Web Conference 2022}{April 25--29, 2022}{Virtual Event, Lyon, France}
\acmBooktitle{Proceedings of the ACM Web Conference 2022 (WWW '22), April
25--29, 2022, Virtual Event, Lyon, France}
\acmPrice{15.00}
\acmDOI{10.1145/3485447.3512164}
\acmISBN{978-1-4503-9096-5/22/04}

\usepackage{booktabs}
\usepackage{amsmath}
\usepackage{algorithm}
\usepackage{algpseudocode}
\usepackage{multirow}
\usepackage{mathtools}
\usepackage{subfigure}
\usepackage{graphicx}
\usepackage{balance}
\usepackage{color}
\usepackage{enumitem}
\usepackage{xspace}

\renewcommand{\vec}[1]{\ensuremath{\mathbf{#1}}}
\newcommand{\stitle}[1]{\vspace{2mm} \noindent {\bf #1}}
\newcommand{\eg}{{\it e.g.}}

\newcommand{\ie}{{\it i.e.}}

\newcommand{\bV}{\ensuremath{\mathcal{V}}}
\newcommand{\bG}{\ensuremath{\mathcal{G}}}
\newcommand{\bN}{\ensuremath{\mathcal{N}}}
\newcommand{\bE}{\ensuremath{\mathcal{E}}}
\newcommand{\bT}{\ensuremath{\mathcal{T}}}

\newcommand{\bH}{\ensuremath{\mathcal{H}}}

\newcommand{\bI}{\ensuremath{\mathcal{I}}}

\newcommand{\model}{TREND}

\begin{document}

% \fancyhead{}

%% The "title" command has an optional parameter,
%% allowing the author to define a "short title" to be used in page headers.
\title{\model: TempoRal Event and Node Dynamics\\for Graph Representation Learning}

\author{Zhihao Wen}
\affiliation{%
    % \institution{School of Computing and Information Systems, Singapore Management University}
  \institution{Singapore Management University}
   \country{Singapore}
   }
\email{zhwen.2019@smu.edu.sg}

\author{Yuan Fang}
\affiliation{%
    % \institution{School of Computing and Information Systems}
  \institution{Singapore Management University}
  \country{Singapore}
  }
\email{yfang@smu.edu.sg}

% \author{Anonymous}
%%
%% By default, the full list of authors will be used in the page
%% headers. Often, this list is too long, and will overlap
%% other information printed in the page headers. This command allows
%% the author to define a more concise list
%% of authors' names for this purpose.
\renewcommand{\shortauthors}{Zhihao Wen and Yuan Fang}

%%
%% The abstract is a short summary of the work to be presented in the
%% article.
\begin{abstract}
Temporal graph representation learning has drawn significant attention for the prevalence of temporal graphs in the real world. However, most existing works resort to taking discrete snapshots of the temporal graph, or are not inductive to deal with new nodes, or do not model the exciting effects which is the ability of events to influence the occurrence of another event. In this work, We propose \model, a novel framework for temporal graph representation learning, driven by \underline{T}empo\underline{R}al \underline{E}vent and \underline{N}ode \underline{D}ynamics and built upon a Hawkes process-based graph neural network (GNN). \model\  presents a few major advantages: (1) it is inductive due to its GNN architecture; (2) it captures the exciting effects between events by the adoption of the Hawkes process; (3) as our main novelty, it captures the individual and collective characteristics of events by integrating both event and node dynamics, driving a more precise modeling of the temporal process.
Extensive experiments on four real-world datasets demonstrate the effectiveness of our proposed model.
% However, most existing works focus on static graphs or snapshots of temporal graphs while the temporal formation process of graphs has yet seldom been explored.
%  When learning over temporal graphs, there are two fundamental questions:(1) How to effectively model temporal formation process over graphs? (2) How to elegantly encode temporal graph information into proper representations? Hence, in this paper, we propose a novel temporal graph representation method, called TREND, not only being aware of the event-level temporal formation, but also being able to recognize the node-level graph dynamics. More specifically, we learn an  edge-level auxiliary model to make edge-level encoder to be customized to each edge and enable link prediction be more robust. 
\end{abstract}

%%
%% The code below is generated by the tool at http://dl.acm.org/ccs.cfm.
%% Please copy and paste the code instead of the example below.
%%
% \begin{CCSXML}
% <ccs2012>
%   <concept>
%       <concept_id>10010147.10010257.10010293.10010319</concept_id>
%       <concept_desc>Computing methodologies~Learning latent representations</concept_desc>
%       <concept_significance>500</concept_significance>
%       </concept>
%   <concept>
%       <concept_id>10002951.10003227.10003351</concept_id>
%       <concept_desc>Information systems~Data mining</concept_desc>
%       <concept_significance>500</concept_significance>
%       </concept>
%  </ccs2012>
% \end{CCSXML}
% \ccsdesc[500]{Computing methodologies~Learning latent representations}
% \ccsdesc[500]{Information systems~Data mining}



%%
%% Keywords. The author(s) should pick words that accurately describe
%% the work being presented. Separate the keywords with commas.

%% A "teaser" image appears between the author and affiliation
%% information and the body of the document, and typically spans the
%% page.
% \begin{teaserfigure}
%   \includegraphics[width=\textwidth]{sampleteaser}
%   \caption{Seattle Mariners at Spring Training, 2010.}
%   \Description{Enjoying the baseball game from the third-base
%   seats. Ichiro Suzuki preparing to bat.}
%   \label{fig:teaser}
% \end{teaserfigure}

%%
%% This command processes the author and affiliation and title
%% information and builds the first part of the formatted document.

\maketitle

\section{Introduction}\label{sec:intro}
%\fang{before submission: remember to add some relevance to WWW. check the CFP. }
Graph-structured data widely exist in real-world scenarios, \eg, social networks, citation networks, e-commerce networks, and the World Wide Web.
To discover insights from these data, graph representation learning has emerged as a key enabler which can encode graph structures into a low-dimensional latent space.
%keystone way of analysing graph-structured data, due to its capability of encoding the structures and properties of graph with low-dimensional latent representations . 
The state of the art has made important progress and many approaches are proposed, which can be mainly divided into two categories: network embedding \cite{cai2018comprehensive} and graph neural networks (GNN) \cite{wu2020comprehensive}. Network embedding approaches are often transductive, which directly learn node embedding vectors using various local structures, like random walks in DeepWalk \cite{perozzi2014deepwalk} and node2vec \cite{grover2016node2vec}, and 1st- and 2nd-order proximity in LINE \cite{tang2015line}. In contrast, GNNs do not directly learn node embedding vectors. They instead learn an inductive aggregation function \cite{kipf2016semi, hamilton2017inductive, velic2018graph, xu2018powerful,wu2019simplifying} which can be generalized to unseen nodes or even new graphs in the same feature space. Typical GNNs follow a message passing framework, where each node receives and aggregates messages (\ie, node features or embeddings) from its neighboring nodes recursively in multiple layers. In other words, GNNs are capable of not only encoding graph structures, but also preserving node features.
%. In constrast, classical network embedding methods only use the local structural information, which is another major drawback of network embedding.

\begin{figure}[t]
%   \centering
  %\includegraphics[width=0.85\linewidth]{figures/fig-motivation.pdf} 
    \includegraphics[scale=0.5]{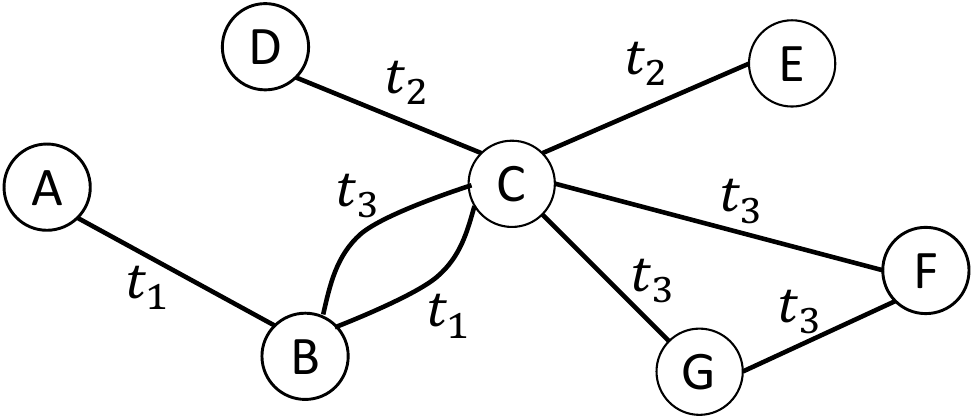}
    \vspace{-2mm}
  \caption{Toy temporal graph for research collaborations that evolves through time $ t_1,t_2, t_3,\cdots $. Each node is a researcher and each link is a collaboration between researchers formed at a specific time. 
  }
  \label{fig:toy node}
  \vspace{-2mm}
\end{figure}

Most of these graph representation methods focus on static graphs with structures frozen in time. However, real-world graphs often present complex dynamics that evolve continuously in time. 
%For instance, social messaging networks, citation networks and e-commerce purchasing networks contain fine-grained temporal influence on edges and nodes that characterize the dynamic evolution of a graph over time. 
For instance, in social networks, burst events often rapidly change the short-term social interaction pattern, while on an e-commerce user-item graph, long-term user preferences may drift as new generations of product emerge.  
%\fang{add a concrete example to illustrate the dynamics in one network. e.g. in social network, how users / relationships ... }. 
More precisely, in a temporal graph \cite{li2017fundamental}, the temporal evolution arises from the chronological formation of links between nodes. As illustrated in Fig.~\ref{fig:toy node}, the toy graph evolving from time $t_1$ through $t_3$     
%from $t_{1}$ to $t_{2}$, the temporal graph structural formation 
can be described by a series of triple $\{(A,B,t_{1}), (B,C,t_{1}), (C,D,t_{2}), (C,E,t_{2}), (B,C,t_{3}), 
\ldots\}$, where each triple $(i, j, t)$ denotes a link formed between nodes $i$ and $j$ at time $t$.
%\fang{do not use $\to$? this looks like two snapshots.}
%The sequence of temporal links encodes network dynamics. 
Hence, the prediction of future links depends heavily on the dynamics embodied in the historical link formation \cite{sarkar2012nonparametric}. 
%reflects the capability of a model for temporal representation learning.
In this paper, we investigate the important problem of \emph{temporal graph representation learning}, in which we learn representations that evolve with time on a graph. In particular, we treat the formation of a link at a specific time as an \emph{event}, and a graph evolves or grows continuously as more events are accumulated \cite{holme2012temporal}.  
%of temporal networks is of great practical application value, for it can help to uncover both the network structure and temporal evolution information hidden in the temporal networks.

\stitle{Prior work.} 
In general, the requirement for temporal graph representation learning is that the learned representations must not only preserve graph structures and node features, but also reflect its topological evolution. However, this goal is non-trivial, and it is not until recently that several works on this problem have emerged. Among them, some \cite{du2018dynamic, li2018deep, goyal2018dyngem, zhou2018dynamic, sankar2020dysat, pareja2020evolvegcn} discretize the temporal graph into a sequence of static graph snapshots to simplify the model. As a consequence, they cannot fully capture the continuously evolving dynamics, for the fine-grained link formation events ``in-between'' the snapshots are inevitably lost.
% As for continuous-time methods, they can be mainly divided into two categories: non-point process  based methods and approaches based on temporal point process \cite{hawkes1971spectra, mei2017neural}, which models sequential events by assuming that historical events before time $t$ can influence the current event with time decay effect. Such property is desirable for modeling the temporal link formation sequences, because the current link formation can be affected with higher intensity by the more recent events, while the events occurring long before would have less influence on the current event formation.
For continuous-time methods, CTDNE \cite{nguyen2018continuous} resort to temporal random walks that respect the chronological sequence of the edges; %proposed continuous-time dynamic network embedding by using a kind of simple time-respect random walks. 
TGAT \cite{xu2020inductive} employs a GNN framework with functional time encoding to map continuous time and self-attention to aggregate temporal-topological neighborhood. %, as well as encodes time as apart of the temporal node representations 
%\fang{non-point process does not sound like a category. Can you summarize these methods with a better name? Also, what are their problems? Why not as good as temporal process}. 
However, these methods often fail to explicitly capture the \emph{exciting effects} \cite{zuo2018embedding} between sequential events, particularly the influence of historical events on the current events.
% , across the entire graph. 
Nonetheless, such effects can be well captured by temporal point processes, most notably the \emph{Hawkes process} \cite{hawkes1971spectra, mei2017neural}, which assumes that historical events prior to time $t$ can excite the process in the sense that future events become more probable for some time after $t$. This property is desirable for modeling the graph-wide link formation process, in which each link formation is considered an \emph{event} that can be excited by recent events. For example, in social networks, a celebrity who has attracted a large crowd of followers lately (\eg, due to winning a prestigious award) is likely to attract more followers in the near future.  
%because the current link formation can be affected with higher intensity by the more recent events, while the events occurring long before would have less influence on the current event formation.
However, for temporal graph representation learning, existing Hawkes process-based network embedding methods \cite{zuo2018embedding, lu2019temporal} are inherently transductive. 
% (\ie, unable to generalize to new nodes during testing, save some na\"ive workarounds). %, and do not benefit from the message passing mechanism in GNNs.
% Meanwhile, high-throughput machine learning systems in real world requires that node representations have inductive capability \cite{hamilton2017inductive} enabling them to be quickly generated for unseen nodes or entirely new graphs in the same domain.
While DyRep \cite{trivedi2019dyrep} presents an inductive framework based on the temporal point process, it addresses a different problem setting of two-time scale with both association and communication events. In our paper, we focus on learning the dynamics of evolving topology, where each event represents the formation of a link.
%learning aggregation functions and temporally parameterizing an adjacent attention matrix,  instead of directly learning node embedding vectors. 
% But, DyRep uses point process to temporally parameterize an adjacent attention matrix, which is not scalable on large graphs. While many temporal graphs in real world are very huge.
%But, DyRep is designed to deal with association and communication events \cite{trivedi2019dyrep}.
%\fang{DyRep: any downside with this problem? e.g. they have a different focus/motivation than our method?}

\stitle{Challenges and present work.}
To effectively model the events of link formation on a temporal graph, we propose a Hawkes process-based GNN framework to reap the benefits of both worlds. Previous methods do not employ Hawkes or similar point processes for modeling the exciting effects between events \cite{xu2020inductive}, or not use message-passing GNNs for preserving the structures and features of nodes in an inductive manner \cite{zuo2018embedding,lu2019temporal}, or neither \cite{nguyen2018continuous}. More importantly, while the Hawkes process is well suited for modeling the graph-wide link formation process, prior methods fail to examine two open challenges on modeling the events (\ie, link formation), as follows.

\textsc{Challenge 1}: \emph{How do we capture the uniqueness of the events on an individual scale?} Different links are often formed out of different contexts and time periods, causing subtle differences among events.  
Taking the research collaboration network in Fig.~\ref{fig:toy node} as an example, while links are all collaborations, each collaboration can be unique in its own way. For instance, the collaboration between researchers $A$ and $B$, and that between $F$ and $G$, could be formed due to  different backgrounds and reasons (\eg, they might simply be students of the same advisor who devised the solution together, or they possess  complementary skills required in a large multi-disciplinary project). Multiple collaborations can also be formed at different times, such as between $B$ and $C$ at $t_1$ and $t_3$, for potentially different reasons. 
%Even for two fixed scholars $B$ and $C$, they form different link with each other at different time.
Conventional methods on temporal graphs train one model to fit all events, where different events tend to pull the model in many opposing directions. The resulting model would be overly diffuse with its center of mass around the most frequent patterns among the events, whilst neglecting many long-tailed patterns covering their individual characteristics. 
%The one-model-fits-all approach cannot sufficiently capture the individual characteristics of events. 
%\emph{how to train a model that can deal with the diversity of temporal events and inter-differences?} 
%While the existing temporal methods \cite{trivedi2019dyrep, xu2020inductive} are inductive, they finally train one inductive model applied to all temporal events. 
%Thus, the one-model-fits-all approach cannot sufficiently capture the individual characteristics of events. 
%turns out to be a major drawback, for neglecting the diversity of temporal events and the inter-event differences. In a temporal social network of real word, events of temporal link formation tend to have large diversity, and the inter-event differences is not negligible.  
Hence, in this paper, motivated by hypernetworks \cite{ha2016hypernetworks,perez2018film, wen2021meta}, we learn an \emph{event prior}, which only encodes the general knowledge of link formation. This event prior can be further specialized in an event-wise manner to fit the individual characteristics of events. 
% While meta-learning has been successfully applied in various kinds of data including texts \cite{hu2019few}, images \cite{liu2019learning} and graphs \cite{zhou2019meta}, prior works mainly deal with few-shot learning. In contrast, our work is the first attempt to learn the \emph{event dynamics} based on an adaptable form of event prior for temporal graph representation learning.

\textsc{Challenge 2}: \emph{How do we govern the occurrence of events on a collective scale?} While events exhibit individual characteristics, they are not formed in isolation and related events often manifest collective characteristics. 
% Given that links are formed to connect nodes, events are inherently related via their common nodes. In particular, different nodes have varying tendency to form new links with others, and even the same node would show different desire to form links at different times. 
Events sharing a common node can be "constrained as a collection" due to the common influence from their shared node. That is, the collection of events of each node should match the arrival rate of the node, which we call node dynamics. Of course, for two different nodes, each would have its own event collection, and each collection should match different arrival rates of the two nodes.
As shown in Fig.~\ref{fig:toy node}, researchers $A$ and $C$ have different tendency to form a collaboration with others at time $t_3$, with $C$ being more active in seeking collaborations. Moreover, as the graph evolves from time $t_{1}$ to $t_{3}$, researcher $C$'s tendency in collaborating with others also evolves to become higher (\eg, due to $C$'s growing reputation). 
In other words, the events stemming from a common node are collectively governed by the dynamics of the node as a function of time.
Hence, we formulate the notion of \emph{node dynamics} to model the collective characteristics of the events from the same node. 
%has different will to form new links and thus form different number of new links, \ie, $\Delta e_{C}(t_{1})=1$ while $\Delta e_{C}(t_{2})=4$. 
%This kind of node dynamic is actually a kind of significant inner driving force of new link formation and whole graph evolution. 
Intuitively, integrating the node dynamics provides a regularizing mechanism beyond individual events, to ensure that the events from a node, as a collection, conform to the continuous evolution of the node.
%significant evolutionary information to enhance the capability of node representation capturing network structure and evolution pattern. Thus, encoding the node-dynamics should be a requirement not to be ignored for temporal graph representation methods. 
Despite the importance of node dynamics, it has not been explored in temporal graph representation learning.

% \stitle{Present work and challenges.}
% Meanwhile, high-throughput machine learning systems in real world requires that node representations have inductive capability \cite{hamilton2017inductive} enabling them to be quickly generated for unseen nodes or entirely new graphs in the same domain. And we notice that, by learning aggregation functions which integrates node features and structures into representation, instead of directly parameterizing node embeddings, GNNs have inherent inductive power. Recently, several inductive GNNs-based temporal representation approaches are proposed \cite{trivedi2019dyrep, xu2020inductive}. However, these works have three fatal shortcomings: (1) When modeling temporal event formation, they just focus on general difference of two nodes related to this edge. Instead, besides temporal GNNs, we build an \emph{transfer function} $\theta_{e}$ which cares about the element-wise difference of two node embeddings; (2) They train a one-model-fits-all approach, neglecting the diversity of edges and ignoring the inter-edge differences that can be crucial to new edges. Meanwhile, we train an \emph{event prior} $\phi$, transforming $\theta_{e}$ into $\theta_{e}^{i, j}$ conditioned on each event; (3) They didn't take the node-dynamics into account. In our work, we train a node-dynamics predictor, which can equip the temporal node representation with node-dynamics information.

\stitle{Contributions.} 
We propose \model, a novel framework for temporal graph representation learning driven by \underline{T}empo\underline{R}al \underline{E}vent and \underline{N}ode \underline{D}ynamics. 
\model\ is built upon a Hawkes process-based GNN, and presents a few major advantages. Firstly, owning to its GNN architecture, it is inductive in nature, \ie, able to handle new nodes at test time.
Secondly, owning to the adoption of the Hawkes process, it maps the graph-wide link formation process to capture a holistic view of the temporal evolution. 
Thirdly, \model\ integrates both the event and node dynamics to respectively capture the individual and collective characteristics of events, which drives a more precise modeling of the link formation process.

In summary, our work encompasses the following contributions. 
% (1) For the first time, we study the problem of integrating the event-dynamics and node-dynamics into inductive temporal graph representation learning; 
(1) For the first time in temporal graph representation learning, we recognize the importance of modeling the events at an individual and collective scale, and formulate them as event and node dynamics.
(2) We propose a novel framework called \model\ with both event and node dynamics to more precisely model events under a Hawkes process-based GNN. On one hand, the event dynamics learns an adaptable event prior to capture the uniqueness of events individually. On the other hand, the node dynamics regularizes the events at the node level to govern their occurrences collectively.
(3) We conduct extensive experiments on four real-world datasets, which demonstrate the advantages of \model.
%\end{itemize}
%(1) 
%(2) We propose a novel temporal graph representation method (TREND). At event level, with a GNN-based temporal point process, it models the link formation process with customized transfer function for each event, and it constrains the node representation being aware of node-dynamics, at node level.; (3) 

\section{Related Work}

% \stitle{Static graph representation learning.} 
% Recently, a large body of graph representation learning  methods has been proposed, including network embedding and graph neural networks (GNNs).
% On one hand, network embedding approaches typically parameterize the node representations directly using an embedding lookup. They can employ shallow models such as the random walk-based skip-gram models \cite{perozzi2014deepwalk, grover2016node2vec} and proximity models \cite{tang2015line}, as well as deep models such as autoencoders \cite{wang2016structural,zhang2018anrl} 
% and adversarial networks \cite{wang2018graphgan}.
% On the other hand, GNNs \cite{wu2020comprehensive} follow a message passing scheme, which employs a learnable aggregation function to map and aggregates messages from neighboring nodes. Different GNNs vary in their choice of the aggregation function, ranging from simple pooling \cite{kipf2016semi,hamilton2017inductive} to the self-attention mechanism \cite{velic2018graph}. However, these methods only deal with static graphs and learn graph representations at a single point of time.

Recently, a large body of graph representation learning  methods has been proposed, including network embedding \cite{cai2018comprehensive} and graph neural networks \cite{wu2020comprehensive}.
To address real-world scenarios in which graphs continuously evolve in time, there have been some efforts in temporal graph representation learning. 
%They mainly belong to two categories: modeling the temporal graph as a series of discrete snapshots, and modeling the continuous process of temporal evolution. 
Intuitively, a temporal graph can be modeled as a series of  snapshots. The general idea is to learn node representations for each graph snapshot, and then capture both the graph structures in each snapshot and the sequential effect across the snapshots. 
The specific techniques vary in different works, 
such as matrix perturbation theory  \cite{li2017attributed, zhu2018high}, skip-grams \cite{du2018dynamic} and triadic closure process \cite{zhou2018dynamic}. 
%Specifically, approaches of  \cite{li2017attributed, zhu2018high} are based on perturbation theory. 
%DNE \cite{du2018dynamic} extend skip-gram based network embedding into dynamic network embedding. 
%DynamicTriad \cite{zhou2018dynamic} leverages the triadic closure process to capture dynamics at each time step. 
To effectively capture the sequential effect, recurrent neural networks (RNNs) have been a popular tool \cite{ijcai2019-640,goyal2020dyngraph2vec,kumar2019predicting,hajiramezanali2019variational},
which leverage the chronological sequence of representations across all snapshots.
%tNodeEmbed \cite{ijcai2019-640} learns the evolution of a temporal graph in an end-to-end architecture for different prediction tasks. 
From a different perspective, instead of using RNNs to generate node representations, EvolveGCN \cite{pareja2020evolvegcn} uses RNN to evolve GCN parameters.
% to capture the dynamics across the sequences of graph snapshots.
Besides, rather than directly learning the representation, DynGEM \cite{goyal2018dyngem} 
% adopts an incremental strategy to 
incrementally builds the representations of a snapshot from those of the previous snapshot. 
%Not only utilizing the snapshot at $t-1$, DynAERNN \cite{goyal2020dyngraph2vec} uses an encoder to acquire network representations from $t-n$ to $t-1$, passes them into LSTM and finally uses a decoder to get the future network structures. 
%JODIE \cite{kumar2019predicting} applies RNNs to estimate the future node embeddings for bipartite graphs.
%Also being a dynamic graph autoencoder model, by integrating GCN and RNN, VGRNN \cite{hajiramezanali2019variational} develops a hierarchical variational model which introduces additional latent random variables to jointly train the graph recurrent neural networks. 
%Instead of using RNN to generate node embeddings, EvolveGCN \cite{pareja2020evolvegcn} uses RNN to evolve GCN parameters to capture dynamic information of sequences of graph snapshots. 

However, snapshots are approximations which discretize a continuously evolving graph, inevitably suffering from a fair degree of information loss in the temporal dynamics. 
To overcome this problem, another line of work aims to model the continuous process of temporal graph evolution, usually by treating each event (typically defined as the formation of a link that can occur continuously in time) as an individual training instance.  
%The second type of approaches encode the evolution pattern of graph into latent representations, by treating each temporal link as an individual training instance. 
Among them, some employ temporal random walks to capture the continuous-time network dynamics, including CTDNE \cite{nguyen2018continuous} based on time-respect random walks, and
CAW-N \cite{wang2021inductive} based on causal anonymous walks.
% which helps to retrieve temporal network motifs as proxies for network dynamics. 
%and substitutes the absolute node identities with relative node identities,  to inductively learn node embeddings.
Apart from random walks, GNN-based models have also emerged to deal with continuous time, e.g., TGAT \cite{xu2020inductive}.
% One example is TGAT \cite{xu2020inductive}, which proposes function time encoding to encode continuous time as part of the temporal representation, together with a self-attention mechanism to aggregate  temporal-topological neighborhood features.
%Based on GraphSAGE and GAT, TGAT \cite{xu2020inductive} aggregates temporal-topological neighborhood features using the self-attention mechanism, and encodes time as apart of the temporal node embedding.
%\fang{check this sentence: But that TGAT's sampling strategy requires to store all historical neighbors makes it unscalable and unpractical to deal with high-throughput evolving graphs in real world. Instead, our TREND directly works each link and only requires to memorize some historical neighbors for the two nodes of interest.}
While these methods can deal with a continuously evolving graph, they fail to explicitly model the exciting effects between sequential events holistically on the entire graph. 
%, particularly the influence of his-tory on the current events. 
In view of this, several network embedding methods \cite{zuo2018embedding, lu2019temporal, DBLP:conf/pkdd/JiJFS21} incorporate temporal point processes such as Hawkes process into their models, being capable of modeling the graph-wide formation process.
%There are also several approaches motivated by Hawkes process.
% More specifically, HTNE \cite{zuo2018embedding} models the neighborhood formation sequence, 
% %to parameterize node embeddings. 
% while MMDNE \cite{lu2019temporal} captures micro- and macro-dynamics of graph formation simultaneously. 
Moreover,  DyREP \cite{trivedi2019dyrep} is an inductive GNN-based model that also exploits the temporal point processes.
%takes the representation as a mediation bridging topological evolution and activities between nodes. 

Note that, among existing methods for temporal graph representation learning, those employing an embedding lookup for node representations are usually transductive \cite{du2018dynamic, zhou2018dynamic, ijcai2019-640, nguyen2018continuous, zuo2018embedding, lu2019temporal} and thus unable to directly make predictions on new nodes at a future time. In contrast, GNN-based methods \cite{pareja2020evolvegcn, xu2020inductive, trivedi2019dyrep} are naturally inductive, able to extend to new nodes in the same feature space. However, among them only DyREP \cite{trivedi2019dyrep} leverages temporal point processes, but it is designed to capture association and communication events, which differs from our problem setting to specifically deal with the link formation process.
%some learn node embeddings on discrete snapshots \cite{goyal2018dyngem, goyal2020dyngraph2vec, hajiramezanali2019variational, pareja2020evolvegcn} \fang{citation} 
%and the other model temporal process of graphs with limited dynamics \cite{trivedi2019dyrep}
%and ignoring the diversity of temporal events \cite{xu2020inductive, wang2021inductive}
Furthermore, none of existing methods integrates both event- and node-dynamics to capture the individual and collective characteristics of events, respectively.
%into temporal graph representation learning.

% \stitle{Representation learning on temporal graphs.}

\begin{figure*}
  \centering
  \includegraphics[scale=0.7]{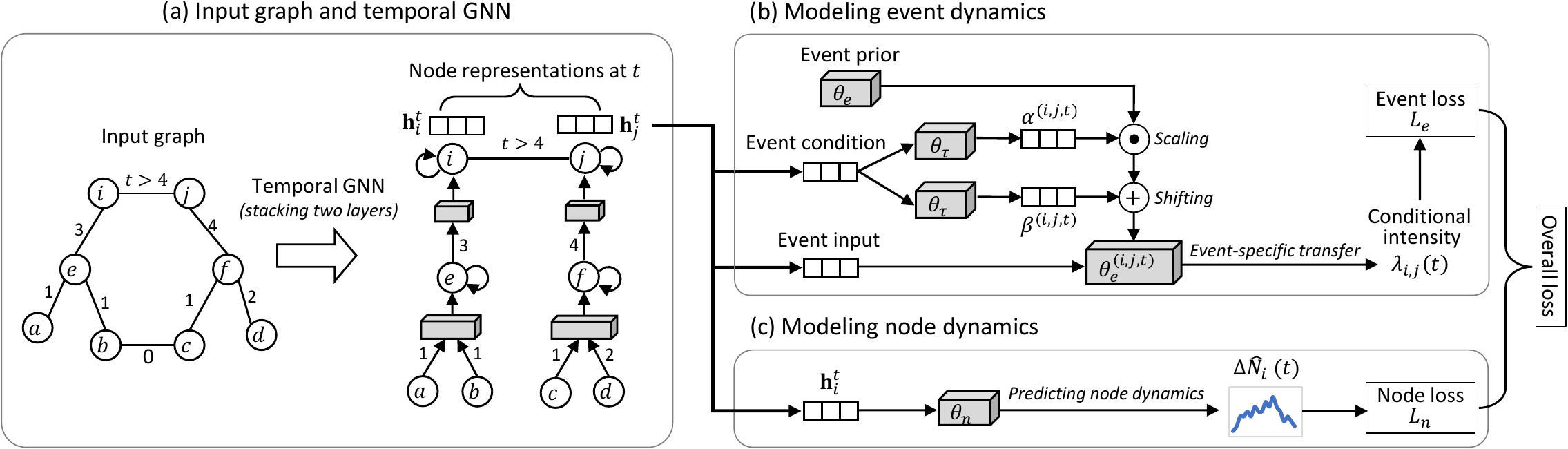}
  \vspace{-1mm}
  \caption{Overall framework of TREND, which integrates event and node dynamics in a Hawkes process-based GNN.}
   \vspace{-1mm}
  \label{fig:framework}
\end{figure*}

\section{PRELIMINARIES}

In this section, we first present the problem of temporal graph representation learning, and then introduce a brief background on the Hawkes process.

\subsection{Temporal Graph Representation Learning}
A \emph{temporal graph} $\bG = (\bV, \bE, \bT, \vec{X})$ is defined on a set of nodes $\bV$, a set of edges $\bE$, a time domain $\bT$ and an input feature matrix $\vec{X} \in \mathbb{R}^{|\bV| \times d_0}$. Each node has a $d_0$-dimensional input feature vector corresponding to one row in $\vec{X}$. 
% An \emph{event} is a triple $(a,b,t)$ denoting the formation of an edge $(a,b)\in \bE$ (also called a link) between node $a\in \bV$ and node $b\in \bV$ at time $t\in \bT$.
% Alternatively, a temporal graph can be defined as a chronological series of events $\{ (a_i,b_i,t_i):i=1,2,3,\ldots \}$.
An \emph{event} is a triple $(i,j,t)$ denoting the formation of an edge $(i,j)\in \bE$ (also called a link) between node $i\in \bV$ and node $j\in \bV$ at time $t\in \bT$.
Alternatively, a temporal graph can be defined as a chronological series of events $ \bI=\left\{(i, j, t)_{m}:m=1,2,\ldots,|\bE|\right\}$.
%A sequence of chronological edges, where each edge links two nodes at a certain time point, forms a temporal graph, which can be denoted as $\bG = (\bV, \bE, \bT)$, where $\bV$ denotes the node set, $\bE$ denotes the edge set, and $\bT$ is the timestamp sequence. A chronological event of edge forming between nodes $i$ and $j$ at time $t$ can be denoted as $(i, j, t)$, the temporal edge.
Note that two nodes may form a link more than once at different times. 
%For example, in a collaboration network, two researchers may collaborate with each other multiple times; in a e-commerce user-item graph, a user may purchase the same item multiple times. 
Thus, there may be two events $(i,j,t_1)$ and $(i,j,t_2)$ such that $t_1\ne t_2$. Besides, in this work, we only consider the growth of temporal graph, and make deletions of node and edge as future work.

We study the problem of inductive temporal graph representation learning. Specifically, given $\bG = (\bV, \bE, \bT, \vec{X})$, we aim to learn a parametric model $\Phi(\cdot;\theta)$ with parameters $\theta$, such that $\Phi$ maps any node in the same feature space of $\vec{X}$ at any time to a representation vector. That is, $\Phi : \bV' \times \bT \to \mathbb{R}^d$, where $\bV \subset \bV'$. The set difference $\bV' \setminus \bV$ consists of new nodes in the same feature space of $\vec{X}$ that may appear at a future time. Such a model $\Phi$ is apparently inductive given the ability to handle new nodes.

\subsection{Hawkes Process}
A Hawkes process \cite{hawkes1971spectra} 
%\fang{use a more classic citation instead of recent embedding paper} 
is a stochastic process that can be understood as counting the number of events up to time $t$. Its behavior is typically modeled by a \emph{conditional intensity} function $\lambda(t)$, the rate of event occurring  at time $t$ given the past events. 
%Formally, in an  infinitesimally small  time window $ [t, t+dt)$, $\lambda(t)dt $ is the conditional probability of an event occurring given all the historical events $\bH(t)$ occurred prior to $t$, \ie, $\lambda(t)dt = \mathbb{P}[event \in [t, t+dt)|\bH(t)] = \mathbb{E}[\mathbf{}{N}(dt)|\bH(t)]$. 
A common formulation of the conditional intensity \cite{zuo2018embedding, lu2019temporal} is given by
%Hawkes process \cite{zuo2018embedding, lu2019temporal} is a typical temporal point process, having the following conditional intensity, 
\begin{align} 
\lambda(t) = \mu(t) + \int_{-\infty}^{t} \kappa (t-s)dN(s),
\label{eq:Hawkes}
\end{align}
where $\mu (t)$ is the base intensity at time $t$, $\kappa$ is a kernel function to model the time decay effect of historical events on the current event (usually in the shape of an exponential function), and $N(t)$ is the number of events occurred until $t$. Since the Hawkes process is able to model the exciting effects between events to capture the influence of historical events holistically, it is well suited for modeling the graph-wide link formation process in a temporal graph.
%Showing that the occurrence of current event does not only be determined by the event of last time step, but also influenced by the historical events with time decay effect \cite{zuo2018embedding}, the conditional intensity function of Hawkes process, is ideal for modeling the temporal event formation sequences. 

\section{Proposed Approach} 
In this section, we present a novel framework for temporal graph representation learning called \model. 
% We start with an overview of our framework, and then introduce its components in detail. 

\subsection{Overview of \model}
%\fang{I moved the following two senences here. They can be apart of framework introduction}
Building upon a Hawkes process-based GNN, the proposed \model\ is able to model the graph-wide link formation process in an inductive manner. More importantly, it integrates event and node dynamics into the model to fully capture the individual and collective characteristics of events.
% Designing an inductive temporal model that not only can deal with the event diversity, but also is able to capture the node dynamics, is the overarching philosophy. 

The overall framework of \model\ is shown in Fig.~\ref{fig:framework}.
First of all, in Fig.~\ref{fig:framework}(a), an input temporal graph undergoes a temporal GNN aggregation in multiple layers, whose output representations serve as the input for modeling event and node dynamics. The GNN layer aggregates both the self-information and historical neighbors' information, which are building blocks to materialize the conditional intensity in the Hawkes process. % (Sect.~\ref{sec:method:event}). 
%we build a Temporal graph neural network(GNN), as shown in Fig.~\ref{fig:framework}(a). Specifically, taking a temporal graph as input, through self propagation and historical local propagation, we can obtain the node representations for modeling the event $(i, j, t)$. 
Next, we model event dynamics to capture the individual characteristics of events, as shown in Fig.~\ref{fig:framework}(b). We perform an event-conditioned, learnable transformation to adapt the event prior to the input event, resulting in an event-specific transfer function to generate the conditional intensity in the Hawkes process. % (Sect.~\ref{sec:customization}). 
%Using the event condition $\vec{h}_{i}^{t}\mid \vec{h}_{j}^{t}$ as input, we can get two transformation factors $\alpha_{i,j}$ and $\beta_{i,j}$, to adjust the transfer function prior $\theta_{e}$ into $\theta_{e}^{i, j}$. Then, given the event representation $(\vec{h}_{i}^{t}-\vec{h}_{j}^{t})^{2}$ as input,  we use the event-specific prior $\theta_{e}^{i, j}$ to compute the conditional intensity $\lambda_{i,j}(t)$.
Moreover, we model node dynamics to capture the collective characteristics of events at the node level, as shown in Fig.~\ref{fig:framework}(c). We build an estimator to  predict the node dynamics across nodes and times, which governs the behavior of events occurring on the same node. %(Sect.~\ref{sec:method:node}).  
%We learn a FCL to predict the dynamics of node $i$ at time $t$, with $\vec{h}_{i}^{t}$ as input. By doing so, the temporal node representation $\vec{h}_{i}^{t}$ will be aware of dynamics at node level.
At last, we integrate the event and node losses to jointly optimize event and node dynamics. % (Sect.~\ref{sec:method:overall}).  
%$L_{e}$ in Fig.~\ref{fig:framework}(b) and node loss $L_{n}$ in Fig.~\ref{fig:framework}(c) together to optimize the whole model.
% Designing an inductive model that can generate node embedding being aware of temporal link formation and new temporal link number of a node, is the overarching philosophy. When talking about designing inductive temporal model, it is easy to think of training a deep model using not only the graph, but also the time information as input. However, how to integrate the graph and time information to maximize the model's ability to predict both the temporal link and temporal link number of a node is still a tough task. Hence, we propose TREND, generating temporal node embedding from historical neighbors, having one link prediction layer being able to customize to different edges, and having one node layer predicting the number of new temporal link of node. 
%\fang{There are still problems with 4.1.\\
%1. each part can be expanded a little more. One sentence is too brief.\\
%2. please link to the framework , i.e., you are describing the figure. but currently i dont see anything that relates the figure.\\
%3. should include references to the parts (a) (b) .. in the text too.}

\subsection{Hawkes process-based GNN}\label{sec:method:event}
We first introduce a Hawkes process-based GNN framework, which is to be further integrated with event and node dynamics later.

\stitle{Hawkes process on temporal graph.}
In the context of temporal graph, the Hawkes process is able to model the graph-wide link formation process. 
%Without loss of generality, we consider directed links such that $(i,j,t)$ and $(j,i,t)$ are two different events. Undirected links are treated as two directed links in opposite directions.
Specifically, whether nodes $i$ and $j$ form a link at $t$, can be quantified by the conditional intensity of the event,
%is influenced by the historical neighbors of the two nodes with a time decay, as follows.
%temporal events of a dynamic graph contain patterns of Hawkes process. At any time $t$, whether node $i$ and node $j$ will form an temporal link,  is influenced by the historical neighbors of the two nodes, with a time decay effect. Thus, it is naturally to apply Hawkes process to model the events of temporal link formation, and the conditional intensity function can be formulated as,
\begin{align}
%\lambda_{i,j}(t)= \mu_{i, j}(t) + \sum_{t_{h}<t} \gamma_{t_{h}} \kappa(t-t_{h})),
\lambda_{i,j}(t)= \mu_{i, j}(t)& + \sum_{(i,j',t') \in \bH_i(t)} \gamma_{j'}(t') \kappa(t-t') \nonumber\\
&+\sum_{(i',j,t') \in \bH_j(t)} \gamma_{i'}(t') \kappa(t-t').
\label{eq:lambda 0}
\end{align}
%\fang{I think $\alpha_{t_{h}}$ should not be here? it is in temporal GNN. and $\kappa$ is typically already exponential} 
% And because $\hat{\lambda}_{i,j}(t)$ is determined by $i$ and $j$ simultaneously, it can be re-write as,
% \begin{align}
% \hat{\lambda}_{i,j}(t)= (\mu_{i} + \sum_{t_{h_{i}}<t} \alpha_{t_{h_{i}}} \kappa(t-t_{h_{i}})) + (\mu_{j} + \sum_{t_{h_{j}}<t} \alpha_{t_{h_{j}}} \kappa(t-t_{h_{j}}))
% \label{eq:lambda 1}
% \end{align} 
%\fang{it should not be separate into i and j. Separated means they are independent. just $\mu_{i,j}$ together, which correspond to i and j's vector input. For the historical event, also just one term, only historical neighbor of i, assume the event is directed from i to j, to have better correspondence with the temporal GNN. I think we can make the directed assumption clear upfront, and state that an undirected link is simply two events in both directions.  }
In particular, $\mu_{i, j}(t)$ is the base rate of the event that $i$ and $j$ form a link at time $t$, which is not influenced by historical events on $i$ or $j$.
$\bH_i(t)$ is the set of historical events on $i$ w.r.t.~time $t$, \ie, $\bH_i(t)=\{ (i,j',t')\in \bI:t'<t\}$, and we call $j'$ a historical neighbor of $i$. $\gamma_{j'}(t')$ represents the amount of excitement induced by a historical neighbor $j'$ at $t'$ on the current event. Note that we are treating each link as undirected, and thus the current event is influenced by historical neighbors of both nodes $i$ and $j$. In the case of directed link, we can modify Eq.~\eqref{eq:lambda 0} by keeping only one of the two summation terms. $\kappa(\cdot)$ is a kernel function to capture the time decay effect w.r.t.~$t$, defined in the form of an exponential function as
%\begin{align}
$\kappa(t-t')=\exp(-\delta(t-t'))$,
%\end{align}
where $\delta > 0$ is a learnable scalar to control the rate of decay.
%\fang{according to preliminary, $\kappa$ itself is already an exponential, make sure that $\delta$ is positive.}

Next, temporal graph representations are used to materialize the conditional intensity in Eq.~\eqref{eq:lambda 0}. 
Given the temporal representations of nodes $i,j$ at time $t$, denoted $\vec{h}_i^t,\vec{h}_j^t$ respectively, the conditional intensity can be generated from a \emph{transfer function}  $f$ \cite{trivedi2019dyrep, mei2017neural}, \ie,
%determined by $i$ and $j$ themselves, representing the base rate of the event of forming a link between $i$ and $j$.  $\gamma_{t_{h}}$ represents the influence of a historical neighbor of $i$ or $j$ to the current event, and the kernel function denotes the time decay effect.
%And it is easy to learn a transfer function \cite{trivedi2019dyrep, mei2017neural} $f_{e}$ to compute $\lambda_{i,j}(t)$, taking $\vec{h}_{i}^{t}$ and $\vec{h}_{j}^{t}$ as input, 
\begin{align}
%\lambda_{i,j}(t)=f_{e}(\vec{h}_{i}^{t}, \vec{h}_{j}^{t}),
\lambda_{i,j}(t)=f(\vec{h}_{i}^{t}, \vec{h}_{j}^{t}), 
\label{eq:lambda 1}
\end{align}
which should meet the following criteria. (1) The input representations $\vec{h}_i^t,\vec{h}_j^t$ should be derived from not only their inherent self-information, but also their historical neighbors' information. While the self-information is the basis of the base intensity $\mu_{i,j}(t)$, historical neighbors are crucial to model the excitement induced by historical events.    (2) The output of the transfer function $f$ must be positive, since it represents an intensity. 
%In previous approaches \cite{trivedi2019dyrep, mei2017neural}, the transfer function is instantiated as a 
%softplus function, which is inflexible in capturing the event dynamics arising from the uniqueness of individual events. Hence, we will employ a learnable transfer function that can be quickly adapted to the events individually, 
We will discuss the choice of the transfer function in Sect.~\ref{sec:customization}.
%where the temporal node representations $\vec{h}_{i}^{t}$ and $\vec{h}_{j}^{t}$ should contains two kinds of information: the inner property of themselves, and historical neighbors' influence with time decay effect. As for the transfer function $f_{e}$, the output should be a positive scalar, since $\lambda_{i,j}(t)$ should take positive value when regarded as a conditional intensity per unit time. And the transfer function of previous methods \cite{mei2017neural, trivedi2019dyrep} are often softplus functions, which is not very flexible to capture the relationship between event representation and event intensity. Hence, we will use a learnable transfer function to better obtain the mapping between event representation and event intensity, as illustrated in Sect.~\ref{sec:customization}.
%\fang{list the criteria for the transfer function as well. e.g. must be positive, and xxx. then, say previous method often use what (softplus?), and state that because of xxx, we will use a learnable function, see later.}

\stitle{Temporal GNN layer.}
We materialize the temporal representations in Eq.~\eqref{eq:lambda 1} using GNNs, owning to their inductive nature and superior performance. 
% \stitle{Temporal graph neural networks.}
%When talking about node representations, we will think of  graph neural networks(GNNs), 
Based on the message passing scheme, each node receives, aggregates and maps messages (\eg, features or embeddings) from its neighboring nodes recursively, in  multiple layers.
%, showing great inductive ability to encode node features and local structures information. 
%Thus, we refer to the Hawkes process and construct the temporal GNNs, sepcifically, in each layer, 
Here we present a temporal formulation of GNN in consideration of the representational criteria listed above, so that the learned temporal representations can be used to materialize the conditional intensity. Formally, let  $\vec{h}_{i}^{t, l} \in \mathbb{R}^{d_{l}}$ be the $d_{l}$-dimensional embedding vector of node $i$ at time $t$ in the $l$-th layer, which is computed by
\begin{align}
\hspace{-1mm}\vec{h}_{i}^{t, l}=\sigma\Big(\hspace{-1mm}\underbrace{\vec{h}_{i}^{t, l-1} \vec{W}_\text{self}^{l}}_\text{\parbox{2cm}{\centering self-information\\[-4pt](for base intensity)}}+\underbrace{\sum_{(i,j',t') \in \bH_{i}(t)}  \vec{h}_{j'}^{t', l-1}\vec{W}_\text{hist}^{l}\tilde{\kappa}_i(t-t')}_\text{\parbox{3.8cm}{\centering historical neighbors' information\\[-4pt](for excitement by historical events)}}\hspace{-.5mm}\Big),\hspace{-1mm}
\label{eq:tgnn}
\end{align}
where $\sigma$ is an activation function (\eg, ReLU), $\vec{W}_\text{self}^{l} \in \mathbb{R}^{d_{l-1}\times d_{l}}$ is a learnable weight matrix to map the embedding of node $i$ itself from the previous layer, % into a $d_{l}$-dimensional vector,
$\vec{W}_\text{hist}^{l} \in \mathbb{R}^{d_{l-1}\times d_{l}}$ is another learnable weight matrix to map the embeddings of historical neighbors, and $\tilde{\kappa}_i(t-t')$ captures the time decay effect based on the time kernel with softmax, which is given by
%\begin{align}
$\tilde{\kappa}_i(t-t')=\frac{\kappa(t-t')}{\sum_{(i,j'',t'')\in\bH_i(t)} \kappa(t-t'')}$.
%\end{align}

In other words, the temporal representation of a node is derived by receiving and aggregating messages of itself and the historical neighbors from the previous layer.
The self-information is responsible for capturing the base intensity, while the historical neighbors' information is responsible for capturing the excitement induced by historical events.
To enhance the representational capacity, we stack multiple temporal GNN layers. In the first layer, the node message can be initialized by the input node features $\vec{X}$;
in the last layer, the output temporal representation is denoted as $\vec{h}_{i}^{t}\in \mathbb{R}^d$ for node $i$ at time $t$.
The collection of parameters of all the layers is $\theta_g=\{\vec{W}_\text{self}^{l}, \vec{W}_\text{hist}^{l}:l=1,2,\ldots\}$.

\stitle{Connection to conditional intensity.}
A well chosen transfer function $f$, taking the temporal representations as input, is equivalent to the conditional intensity of the Hawkes process in Eq.~\eqref{eq:lambda 0}. We formally show the connection in Appendix~\ref{app:proof}.

\subsection{Modeling Event Dynamics}\label{sec:customization}
%\fang{event dynamics -- consistent with the contribution/challenge}
% \stitle{transfer function.}

%As established in Appendix~\ref{app:proof}, %Eq.~\eqref{eq:lambda_equivalence}, 
The key to materialize the conditional intensity is to fit a transfer function $f$ on top of the temporal GNN layers.
Previous studies on Hawkes process employ the softplus function or its variant \cite{mei2017neural, trivedi2019dyrep} as the transfer function. To ensure that $f$ is well-fit to the conditional intensity, our first proposal is to instantiate $f$ as a learnable function. 
More specifically, we use a fully connected layer (FCL). That is,
\begin{align}
\lambda_{i,j}(t)= f(\vec{h}_{i}^{t}, \vec{h}_{j}^{t}) = \operatorname{FCL}_e((\vec{h}_{i}^{t}-\vec{h}_{j}^{t})^{\circ 2};\theta_{e}),
\label{eq:transfer function}
\end{align}
where $\theta_e$ denotes the parameters of the fully connected layer FCL$_e$.
%As for that how to get a conditional intensity,  the existing approach \cite{trivedi2019dyrep} uses a kind of modified softplus function as transfer function. While we train a FCL as the transfer function, which is more flexible than the softplus function. 
Note that the input to FCL$_e$ can be in various forms, such as the concatenation of $\vec{h}_{i}^{t}$ and $\vec{h}_{j}^{t}$, or the element-wise square (denoted by $^\circ 2$) of the difference between them. We use the latter in our formulation, which tends to achieve better empirical performance. A potential reason is that the differential representation is a good predictor of whether an event occurs between the two nodes. Lastly, FCL$_e$ employs a sigmoid activation, to ensure the transfer function is positive.  
%And the output activation function $\sigma$ of transfer function is a sigmoid function, to ensure that the event occurrence intensity is a non-negative probability scalar.
%the transfer function is as follows,
% Then, instead of mimicing input (concatenation of two node representations) of the modified softplus function \cite{trivedi2019dyrep}, we take the squared element-wise representation difference (which can better show the difference or similarity of two nodes, hence better represent the edge) as input to model the occurrence intensity of the event $(i, j, t)$, 
%\fang{I think it is not needed to explain too much why squared difference is used instead of concatenation. Both are possible, there is no convincing argument of one over another. We choose squared difference mainly because empirical results are better. so just be honest--state that we have different choices like x and y, but we find y better empirically and a *potential* reason is it can better capture the difference  better (must say it is potential reason, as it is not a solid black and white answer. }
%\begin{align}
%\lambda_{i,j}(t)= \textsc{FCL}_{e}((\vec{h}_{i}^{t}-\vec{h}_{j}^{t})^{2};\theta_{e}).
%\label{eq:transfer function}
%\end{align}
%where $\theta_{e}$ are learnable matrices of $\textsc{FCL}_{e}$, called event prior. 

Meanwhile, we recognize that each event can be unique in its own way, as different links are often formed out of different contexts and time periods.
To precisely capture the uniqueness of events on an individual scale (\textsc{Challenge 1}), a global  model in Eq.~\eqref{eq:transfer function}---our first proposal---becomes inadequate. To be more specific, in a conventional one-model-fits-all approach, given the diversity in events, the learned model tends to converge around the most frequent patterns among events, while leaving long-tailed patterns that reflect the individual characteristics of events uncovered. On the other hand, training a large number of models for different kinds of events can easily cause overfiting and scalability issues, not to mention that it is difficult to categorize events in the first place. 
Inspired by meta-learning, particularly the line of work on \emph{hypernetworks} \cite{ha2016hypernetworks,perez2018film}, 
we address the dilemma by learning an \emph{event prior}, which can be quickly adapted to a unique model for each event, without the need to train a large number of models.

\stitle{Event prior and adaptation.}
In our first proposal in Eq.~\eqref{eq:transfer function}, we learn a global model for all events, \ie, the same $\theta_e$ parameterizes a global FCL$_e$ as the transfer function for all events.
To deal with the diversity of events, we propose to learn an event prior $\theta_e$ that aims to encode the general knowledge of link formation, such that it can be quickly specialized to fit the individual characteristics of each event. 
In other words, $\theta_e$ does not directly parameterize FCL$_e$ used as the transfer function; instead, it will be adapted to each event via a learnable transformation model first, and the adapted parameters will instantiate an event-specific FCL$_e$ as the transfer function for each event. This approach is a form of hypernetwork \cite{ha2016hypernetworks}, in which a secondary neural network is used to generate the parameters of the primary network. This means the parameters of the primary network can flexibly adapt to its input, as opposed to conventional models whose parameters are frozen once training is completed. In our context, the primary network is FCL$_e$ for the transfer function, and the secondary network is the learnable transformation model.
Particularly, during the adaptation,
%there is a need to adjust the transfer function according to different edges, to better model the event intensity rate $\lambda_{i,j}(t)$. We employ an event-condition prior $\phi$ to condition the event  prior $\theta_{e}$, to make the event  prior customized to each event. 
the event prior $\theta_e$ will transform into event $(i,j,t)$-specific parameters $\theta_e^{(i,j,t)}$ as follows.
%, which acts as the transfer function to materialize the conditional intensity $\lambda_{i,j}(t)$.
\begin{align}
    \theta_e^{(i,j,t)} = \tau (\theta_{e}, \vec{h}_{i}^{t}\| \vec{h}_{j}^{t}; \theta_\tau), \label{eq:tau}
\end{align}
which (1) is parameterized by $\theta_\tau$; (2) is conditioned on event-specific temporal representations of nodes $i,j$, namely, $\vec{h}_{i}^{t}\| \vec{h}_{j}^{t}$ where $\|$ is the concatenation operator; (3) outputs adapted parameters $\theta_{e}^{(i, j,t)}$ by transforming the event prior $\theta_e$ conditioned on $\vec{h}_{i}^{t}\| \vec{h}_{j}^{t}$. The transformed $\theta_{e}^{(i, j,t)}$ will further parameterize FCL$_e$ as the transfer function, and materialize the conditional intensity below. 
\begin{align}
\lambda_{i,j}(t)= \operatorname{FCL_e}((\vec{h}_{i}^{t}-\vec{h}_{j}^{t})^{\circ 2};\theta_{e}^{(i, j,t)}).
\label{eq:lambda}
\end{align}
%That is to say, the event-condition prior does not directly specify the transformation \cite{wen2021meta}, but it encodes the rule of how to transform \wrt each event. This is actually a kind of hypernetwork \cite{ha2016hypernetworks,perez2018film}, in which the event  prior is adjusted by a auxiliary network (parameterized by $\phi$) in response to the diverse input event. 
%inputs the event  prior $\theta_{e}$, and event condition $\vec{h}_{i}^{t}\mid \vec{h}_{j}^{t}$ which is the concatenation of two node representations; (3) outputs a transformed, event-conditioned event  prior $\theta_{e}^{i, j}$. That is to say, the event-condition prior does not directly specify the transformation \cite{wen2021meta}, but it encodes the rule of how to transform \wrt each event. This is actually a kind of hypernetwork \cite{ha2016hypernetworks,perez2018film}, in which the event  prior is adjusted by a auxiliary network (parameterized by $\phi$) in response to the diverse input event. 

In the following, we will materialize the transformation model $\tau$ and its parameters $\theta_\tau$ in detail.

\stitle{Learnable transformation.}
%Conditioned on the event-specific representations $\vec{h}_{i}^{t}\mid \vec{h}_{j}^{t}$, 
We consider Feature-wise Linear Modulation (FiLM) \cite{perez2018film}, which employs affine transformations including scaling and shifting on the event prior, conditioned on event-specific temporal representations.
%we perform feature-wise linear modulations \cite{perez2018film} on the event  prior $\theta_{e}$ to achieve adapting to this event. 
Compared with gating \cite{xu2015show} which can only adjust the parameters in a diminishing way, FiLM is more flexible in adjusting the parameters and can be conditioned on  arbitrary input. Specifically, we employ fully connected layers to generate the scaling operator $\alpha^{(i,j,t)}$ and shifting operator $\beta^{(i,j,t)}$, conditioned on the event-specific input $\vec{h}_{i}^{t}\| \vec{h}_{j}^{t}$, as follows. 
\begin{align}
\alpha^{(i,j,t)}=\sigma\big((\vec{h}_{i}^{t}\| \vec{h}_{j}^{t}) \vec{W}_{\alpha} + \vec{b}_{\alpha}\big)
,\label{eq:alpha}\\
\beta^{(i,j,t)}=\sigma\big((\vec{h}_{i}^{t}\| \vec{h}_{j}^{t})\vec{W}_{\beta} + \vec{b}_{\beta}\big), \label{eq:beta}
\end{align} %\fang{consistent with conditional intensity, use i,j with comma}
where $\vec{W}_{\alpha}, \vec{W}_{\beta} \in \mathbb{R}^{2d \times  d_{\theta_{e}}}$ and $\vec{b}_{\alpha},\vec{b}_{\beta} \in \mathbb{R}^{d_{\theta_{e}}}$ are learnable weight matrices and bias vectors of the fully connected layers, in which $d$ is the dimension of node representation and $d_{\theta_{e}}$ is the total number of parameters in the event  prior $\theta_{e}$. 
%$\vec{b}_{\alpha} \in \mathbb{R}^{d_{\theta_{e}}}$ and $\vec{b}_{\beta} \in \mathbb{R}^{d_{\theta_{e}}}$ are two learnable bias vectors.
The output $\alpha^{(i,j,t)},\beta^{(i,j,t)} \in \mathbb{R}^{d_{\theta_{e}}}$ are both $d_{\theta_{e}}$-dimensional vectors, which represent the scaling and shifting operations of the transformation model $\tau$. %in Eq.~\eqref{eq:tau}. 
They are used to transform the event prior into event $(i, j, t)$-specific parameters by element-wise scaling and shifting, given by 
\begin{align}
\theta_{e}^{(i, j,t)}=\tau (\theta_{e}, \vec{h}_{i}^{t}\| \vec{h}_{j}^{t}; \theta_\tau)=(\alpha^{(i,j,t)} + \vec{1})\odot \theta_{e} + \beta^{(i,j,t)},
\label{eq:event transform}
\end{align}
where $\odot$ stands for element-wise multiplication, and $\vec{1}$ is a vector of ones to ensure that the scaling factors are centered around one.
Note that $\theta_{e}$ contains all the weights and biases of FCL$_e$, and we flatten it into a $d_{\theta_{e}}$-dimensional vector.
%in a slight abuse of notation for brevity.

%Given that $\theta_{e}$ has the same dimension as $\alpha_{i,j}$ and $\beta_{i,j}$, we are able to the conduct the transformation in an element-wise manner, to generate the event-conditioned event  prior,
%\begin{align}
%\theta_{e}^{i, j}=\tau (\theta_{e}, \vec{h}_{i}^{t}\mid \vec{h}_{j}^{t}; \phi)=(\alpha_{i,j} + \vec{1})\odot \theta_{e} + \beta_{i,j},
%\label{eq:event transform}
%\end{align} \fang{better to indicate i,j instead of prime. e.g. $\theta_{e}^{i,j}$}
%where $\vec{1}$ is a vector of ones to ensure that the scaling factors are centered around one, and $\odot$ denotes element-wise multiplication. To note that the event-condition prior $\phi$, forming the parameters of $\tau$, is composed of the parameters of the two set of matrix and bias vector, \ie, 
In summary, the learnable transformation model $\tau$ is parameterized by $\theta_\tau=\{\vec{W}_{\alpha}, \vec{b}_{\alpha}, \vec{W}_{\beta}, \vec{b}_{\beta}\}$, \ie, the collection of parameters of the fully connected layers that generate the scaling and shifting operators. 
%\begin{align}
%    \phi=\{\phi_{\alpha}, \phi_{\beta}\}=\{(\vec{W}_{\alpha}, \vec{b}_{\alpha}), (\vec{W}_{\beta}, \vec{b}_{\beta})\}.
%    \label{eq:phi}
%\end{align}
Furthermore, $\tau$ is also a function of the event condition $\vec{h}_{i}^{t}\| \vec{h}_{j}^{t}$, for $\alpha^{(i,j,t)}$ and $\beta^{(i,j,t)}$ are functions of the event condition.  
%That means $\tau$ is event-specific, adapting to in response to the diverse event. In particular, the two FCLs are actually the secondary networks in the hypernetwork setting \cite{ha2016hypernetworks}.

\stitle{Event loss.}
%Given the event-specific event  prior $\theta_{e}^{i, j}$, we use the aforementioned $\textsc{FCL}_{e}$ to model the occurrence intensity of the event $(i, j, t)$, as follows:
%\begin{align}
%\lambda_{i,j}(t)= \textsc{FCL}_{e}((\vec{h}_{i}^{t}-\vec{h}_{j}^{t})^{2};\theta_{e}^{i, j}).
%\label{eq:lambda}
%\end{align}
% We can see that output of the FCLs is boundless, while the event occurrence intensity is a non-negative probability scalar. Thus, we apply a sigmoid function to transfer the conditional intensity rate to a non-negative probability rate. Finally, we can define the event occurrence intensity as:
% \begin{align}
% \lambda_{i, j}(t) = \sigma( \hat{\lambda}_{i, j}(t)).
% \label{eq:lambda}
% \end{align}\fang{make the sigma apart of the FCL - i.e. the FCL already outputs between 0-1}
Given an event $(i,j,t)\in \bI$ that has occurred on the graph, we expect a higher conditional intensity $\lambda_{i,j}(t)$.
On the contrary, given an event $(i,j,t)\notin \bI$ that does not happen, we expect  a lower conditional intensity.
%a higher \emph{survival rate} $\s_{i,j}(t)$ which is defined as the rate of event not occurring at time $t$, $s_{i,j}(t)=1-\lambda$
%
%Similarly, considering the negative instance, we characterize the rate of event $(i, j^{\prime}, t)$ which does not happen at all, as survival rate \cite{aalen2008survival} $s_{i, j^{\prime}}(t)$,  
%\begin{align}
%s_{i, j^{\prime}}(t) = 1 - \textsc{FCL}_{e}((\vec{h}_{i}^{t}-\vec{h}_{j^{\prime}}^{t})^{2};\theta_{e}^{i, j^{\prime}}).
%\label{eq:survive}
%\end{align}
%
% which is inversely proportional to $\hat{\lambda}_{i, j^{\prime}}(t)$. 
Thus, we formulate the event loss based on negative log-likelihood, the optimization of which encourages the conditional intensity of an event to match its occurrence or non-occurrence.
Given any event $(i,j,t)\in \bI$ that has occurred, its loss is defined as 
\begin{align}
L_{e}(i,j,t)=-\log (\lambda_{i, j}(t))-Q \cdot \mathbb{E}_{k \sim P_n} \log (1-\lambda_{i, k}(t)),
\label{eq:event loss}
\end{align}
where we sample a negative node $k$ according to the distribution $P_n$, so that $(i,k,t)\notin \bI$ does not occur, and $Q$ is the number of negative samples for each positive event. As a common practice, $P_n$ is defined on the node degrees, namely, $P_{n}(v) \propto \operatorname{deg}(v)^\frac{3}{4}$ where $\operatorname{deg}(v)$ is the degree of node $v$. 

\subsection{Modeling Node Dynamics}\label{sec:method:node}
Different from the event dynamics that captures the individual characteristics of events,
node dynamics aims to govern the collective characteristics of events. While events can be individually different, they do not occur in isolation.
Particularly, links are formed to connect nodes, which means their behaviors are collectively bounded by their common nodes.
Thus, we propose to govern the collective characteristics of nodes at the node level, to capture the ``event tendency'' of nodes (\textsc{Challenge 2})---different nodes have varying tendency to form new links with others, and even the same node would manifest different tendency at different times. 

\stitle{Estimator of node dynamics.}
More specifically, the node dynamics or the event tendency of a node at time $t$ can be quantified by the number of new events occurring on the node at $t$, denoted $\Delta N_i(t)$. 
We build an estimator for node dynamics with a fully connected layer, trying to fit the number of new events on a given node:
\begin{align}
\Delta \hat{N}_i(t)= \operatorname{FCL}_{n}(\vec{h}_{i}^{t};\theta_{n}),
\label{eq:node}
\end{align}
where the input is the temporal representation $\vec{h}_{i}^{t}$, the output $\Delta \hat{N}_i(t)$ is the predicted number of new events occurring on node $i$ at time $t$, and $\theta_{n}$ contains the parameters of FCL$_n$.

\stitle{Node loss.}
To ensure that the occurrence of events are consistent with the node dynamics evolving continuously on a temporal graph, we formulate a node loss 
such that the estimator $\Delta \hat{N}_i(t)$ can accurately reflect the groundtruth dynamics $\Delta N_i(t)$ across all nodes and times.
%With the node-dynamics evolving, we have all the ground truth number of new links per node at different time steps. Then, we train the $\textsc{FCL}_{n}$ and enhance our whole model using the following smooth $L_{1}$ loss \cite{girshick2015fast},
In particular, we adopt the following smooth $L_{1}$ loss \cite{girshick2015fast}:
\begin{align}
  \hspace{-1mm}  L_{n}(i,t)= \begin{cases}0.5(\Delta \hat{N}_i(t)-\Delta N_i(t))^{2}, & |\Delta \hat{N}_i(t)-\Delta N_i(t)|<1 \\  |\Delta \hat{N}_i(t)-\Delta N_i(t)|-0.5. & \text { otherwise }\end{cases}\hspace{-1mm}
    \label{eq:node loss}
\end{align}
%where $\Delta N_i(t)$ is the ground truth. 
The smooth $L_{1}$ loss can be viewed as a combination of both $L_{1}$ loss and $L_{2}$ loss. It is less sensitive to outliers than the $L_{2}$ loss when the input is large, and it suffers from less oscillations than the $L_1$ loss when the input is small.
In our scenario, there exist some nodes with a very large number of new links at certain times (\eg, due to burst topics on social networks).
To prevent the models from being overly skewed to these nodes, and to simultaneously cater to nodes with only a few links, the smooth $L_{1}$ loss is an ideal choice. 
%\ie, there are some outliers not to be ignored. So we choose smooth $L_{1}$ loss. And we 
%neither choose $L_{2}$ loss which is very sensitive to these outliers, nor use $L_{1}$ loss which may result in exploding gradients in some cases.
%Intuitively, integrating the node dynamics provides a regularizing mechanism beyond individual events, to ensure that the events occurring on a common node collectively conform to the continuous evolution of the node.

\subsection{Overall Model: \model}\label{sec:method:overall} %\fang{this should be a separate section}.
Finally, we integrate both event and node dynamics into a Hawkes process-based GNN model, resulting in our proposed model \model.
%mutually boosting the evolution of the temporal graph, we get our final model to capture the event formation process and recognize node dynamics in a unified manner. 
Consider the set of training events $\bI^\text{tr}=\{(i,j,t) \in \bI : t\le t^\text{tr}\}$, \ie, all events on the graph up to time $t^\text{tr}$. (New events after time $t^\text{tr}$ can be reserved for testing.)
We optimize all parameters $\Theta=(\theta_{g},\theta_{e},\theta_\tau,\theta_{n})$ jointly, including those of the temporal GNN layers $\theta_{g}$, the event  prior $\theta_{e}$, the transformation model $\theta_\tau$ and the estimator of node dynamics $\theta_{n}$, based on the following loss:
\begin{align}
\arg \min_{\Theta} \sum_{(i, j, t) \in \bI^\text{tr}}L_{e}+ \eta_1 L_{n}+\eta_2(\|\alpha^{(i,j,t)}\|^2_{2}+\|\beta^{(i,j,t)}\|^2_{2}),
\label{eq:overall-optimize}
\end{align}
%\fang{Write the overall loss here? $L_e$ and $L_{n}$ is just a single event/node.} 
where (1) $\eta_1 >0$ is a hyper-parameter controlling the contribution of node dynamics to our model \model; (2) the $L_{2}$ regularization
on $\alpha^{(i,j,t)}$ and $\beta^{(i,j,t)}$
%$\|\alpha_{i,j}\|_{2}+\|\beta_{i,j}\|_{2}$ 
constrains the scaling and shifting operators, as it is preferred that the scaling is close to 1 and the shifting is close to zero, in order to avoid overfitting to individual events; (3) $\eta_2 >0$ is a hyperparameter controlling the effect of the $L_2$ regularizor. 
%$(i, j, t) \in \bI$ , $Q$ negative events $(i, j^{\prime}, t)$, and a temporal link number $\Delta N_i(t)$. Here the goal is to optimize all the priors jointly, including parameters of GNNs, $\theta_{g}$, the event  prior, $\theta_{e}$, the event-condition prior, $\phi$, the node prior, $\theta_{n}$, via backpropagation \wrt the event loss $L_{e}$ and node-dynamics loss $L_{n}$ jointly. Specifically, the optimal $\{\theta_{g}, \theta_{e}, \phi, \theta_{n}\}$ is made by

For implementation, we perform optimization over batches of training events using a gradient-based optimizer. 
%\fang{did not reference algorithm. also , based on the algorithm, offer a complexity analysis.}
The overall training procedure of \model\ is outlined in Appendix~\ref{app:pseudocode}. %Algorithm~\ref{alg:train}. 
It can be seen that the training time complexity is $O(K|\bI^\text{tr}|h^{l}Q))$, where $K$ is the number of epochs, $|\bI^\text{tr}|$ is the number of training events, $h$ is the number of historical neighbors in temporal GNN aggregation, $l$ is the number of temporal GNN layers, and $Q$ is the number of negative samples per training event. Note that $Q$ and $l$ are small constants (typically 5 or less), and $h$ can also be a  constant when employing a commonly used neighborhood sampling approach \cite{hamilton2017inductive}. Hence, the complexity can be regarded as linear in the number of events or temporal edges on the graph.
% And the scalability study is shown in Appendix~\ref{app:scala}. 

\section{EXPERIMENTS}
We conduct extensive experiments to evaluate TREND,
% \footnote{Code and data are available at \textcolor{blue}{\url{https://github.com/666666abc/Trend}} for review.}
with comparison to state-of-the-art baselines and in-depth model analysis.
%To be more specific, TREND is compared with state-of-the art baselines. The effectiveness of transfer function for Hawkes process, event-conditioned transformation and integration of node-dynamics are studied. And furthermore, hyperparameter sensitivity is analyzed.

\subsection{Experimental Setup}
\stitle{Datasets.}
Four public temporal networks are used in our experiments, as summarized in Tab.~\ref{table.datasets}. Note that 
``new nodes in testing'' refers to the ratio of testing events containing at least one new node not seen during training.
(1) \textbf{CollegeMsg} \cite{panzarasa2009patterns}: an online social network in which an event is a user sending another user a private message. (2) \textbf{cit-HepTh} \cite{leskovec2005graphs}: a citation graph about high energy physics theory in which an event is a paper citation.
%We use word2vec\cite{mikolov2013distributed} to convert the text of paper abstraction(the raw node feature) into node embedding as the node feature.
 (3) \textbf{Wikipedia} \cite{kumar2019predicting}: a Wikipedia graph in which an event is a user editing a page. 
 %User edits consist of the textual features, which are converted into 172-dimensional LIWC \cite{pennebaker2001linguistic} feature vectors. The edits vector of each node are added and normalized to serve as the node feature.
    (4) \textbf{Taobao} \cite{du2019sequential}: an e-commerce platform in which an event is a user purchasing an item. %
    %Node features are preprocessed embeddings of textual features.
More dataset details are presented in Appendix~\ref{app:dataset}.

\begin{table}[tbp]
    \small
	\centering  
	%\addtolength{\tabcolsep}{-1pt}
	\caption{Statistics of datasets.}  
	\vspace{-2mm}
	\label{table.datasets}  
	\resizebox{0.99\columnwidth}{!}{
	\begin{tabular}{@{}c|cccc@{}}  
		\toprule  
		Dataset&CollegeMsg&cit-HepTh&Wikipedia&Taobao\\  \midrule
		\# Events &59,835&51,315 &157,474 &4,294,000\\
	%	\# Static edges&20,296&51,315&18,257 &4,048,411\\
		\# Nodes &1,899&7,577&8,227 &1,818,851\\
		\# Node features&$-$&128&172 &128\\
% 		\# Time steps &28&78&31 &36\\
		 Multi-edge? &Yes&No&Yes &Yes\\
		 New nodes in testing &22.79\%&100\%&7.26\% &23.46\%\\

% 		\# Node classes& 7& 10 &7 & 8& 9\\
% 		Multi-label? & No & Yes & Yes & No & No\\
		\bottomrule
	%\bottomrule
	\end{tabular}}
	\vspace{-2mm}
\end{table}

\begin{table*}[tbp]
    % \small%\footnotesize%\scriptsize%%\footnotesize%\small
	\centering 
	\small
 	\addtolength{\tabcolsep}{4pt}
	\caption{Performance of temporal link prediction by TREND and the baselines, in percent, with 95\% confidence intervals.  
	} 
	\label{table:main} 
    {\vspace{-2.5mm}In each column, the best result is \textbf{bolded} and the runner-up is \underline{underlined}. Improvement by TREND is calculated relative to the best baseline. 
    ``-" indicates no result obtained due to out of memory issue; $^{*}$ indicates that our model significantly outperforms the best baseline  based on two-tail $t$-test $(p<0.05)$.
    %\fang{ a bit too many categories. Combine network embedding and GNN. Also, the group with hawkes process put last.}
    }
    \\[2mm] 
	%\resizebox*{\textwidth}{!}{}{
	\begin{tabular}{@{}c|cc|cc|cc|cc@{}}  
		\toprule
		  &\multicolumn{2}{c|}{CollegeMsg}&\multicolumn{2}{c|}{cit-HepTh}&\multicolumn{2}{c|}{Wikipedia}&\multicolumn{2}{c}{Taobao}\\\cmidrule{2-9}
		  & Accuracy&F1&Accuracy&F1&Accuracy&F1&Accuracy&F1
		 \\\midrule
		DeepWalk &66.54$\pm$5.36 &67.86$\pm$5.86 &  51.55$\pm$0.90 & 50.39$\pm$0.98 &65.12$\pm$0.94&64.25$\pm$1.32& 53.59$\pm$0.18 & 56.67$\pm$0.12 \\
		Node2vec &65.82$\pm$4.12&69.10$\pm$3.50&65.68$\pm$1.90&66.13$\pm$2.15&75.52$\pm$0.58&75.61$\pm$0.52&52.74$\pm$0.33&54.86$\pm$0.32 \\
% 		GCN &57.82$\pm$5.68&59.57$\pm$6.14&\underline{76.17}$\pm$1.51&\underline{76.64}$\pm$1.62&&&\underline{68.47}$\pm$0.14&\underline{69.57}$\pm$0.14 \\
        % \midrule
        VGAE &65.82$\pm$5.68&68.73$\pm$4.49&66.79$\pm$2.58&67.27$\pm$2.84&66.35$\pm$1.48&68.04$\pm$1.18&55.97$\pm$0.22&59.80$\pm$0.16 \\
        GAE &62.54$\pm$5.11&66.97$\pm$3.22&69.52$\pm$1.10&70.28$\pm$1.33&68.70$\pm$1.34&69.74$\pm$1.43&58.13$\pm$0.15&61.40$\pm$0.07 \\
		GraphSAGE &58.91$\pm$3.67 &60.45$\pm$4.22 &  70.72$\pm$1.96 & 71.27$\pm$2.41 &72.32$\pm$1.25&73.39$\pm$1.25& 60.74$\pm$0.18 & 61.61$\pm$0.20 \\
		\midrule
		CTDNE &62.55$\pm$3.67 & 65.56$\pm$2.34 &  49.42$\pm$1.86 & 44.23$\pm$3.92 &60.99$\pm$1.26&62.71$\pm$1.49& 51.64$\pm$0.32 & 43.99$\pm$0.38 \\
        EvolveGCN &63.27$\pm$4.42&65.44$\pm$4.72&61.57$\pm$1.53&62.42$\pm$1.54&71.20$\pm$0.88&73.43$\pm$0.51&-&- \\
		GraphSAGE+T  &69.09$\pm$4.91 & 69.41$\pm$5.45 &  67.80$\pm$1.27 & 69.12$\pm$1.12 &57.93$\pm$0.53&63.41$\pm$0.91& 67.05$\pm$0.23 & 67.69$\pm$0.17 \\
		TGAT &58.18$\pm$4.78&57.23$\pm$7.57&\underline{78.02}$\pm$1.93&\underline{78.52}$\pm$1.61&76.45$\pm$0.91&76.99$\pm$1.16&\underline{70.07}$\pm$0.59&\underline{71.31}$\pm$0.18 \\
		\midrule
		HTNE &\underline{73.82}$\pm$5.36 & \underline{74.24}$\pm$5.36 &  66.70$\pm$1.80 & 67.47$\pm$1.16 &77.88$\pm$1.56&78.09$\pm$1.40& 59.03$\pm$0.17 & 60.34$\pm$0.19\\
		MMDNE &\underline{73.82}$\pm$5.36 &74.10$\pm$3.70 &  66.28$\pm$3.87 & 66.70$\pm$3.39 &\underline{79.76}$\pm$0.89&\underline{79.87}$\pm$0.95& 58.24$\pm$0.10 & 59.04$\pm$0.16\\

		\midrule
		TREND & \textbf{74.55}$\pm$1.95 &\textbf{75.64}$\pm$2.09 & \textbf{80.37}$^{*}${}$\pm$2.08 &\textbf{81.13}$^{*}${}$\pm$1.92 &\textbf{83.75}$^{*}${}$\pm$1.19&\textbf{83.86}$^{*}${}$\pm$1.24& \textbf{78.56}$^{*}${}$\pm$0.17 &\textbf{80.67}$^{*}${}$\pm$0.15 \\
		(improv.) & (+0.99\%)&(+1.89\%)& (+3.01\%)&(+3.32\%)&(+5.00\%)&(+4.99\%)& (+12.11\%)&(+13.12\%)\\
% 		(p-value)&(0.84)&(0.58)&($<$0.01) &($<$0.01) &&&($<$0.01) &($<$0.01) \\
% 		\hline 
	\bottomrule
	\end{tabular}%}
	\vspace{-2mm}
\end{table*}

% \stitle{Training and testing.} 
\stitle{Prediction tasks.}
%\fang{here introduce both tasks , including node prediction. (maybe call it "temporal node dynamics prediction"?). State that the main task is link prediction. } 
We adopt \emph{temporal link prediction} as our main task. We evaluate a model by predicting future links based on historical links \cite{sarkar2012nonparametric}.
%reflects the capability of a temporal representation learning model. In our work, we also use temporal link prediction as the main task. 
Given a temporal graph, we split the events into training and testing. Specifically, the set of training events $\bI^\text{tr}=\{(i,j,t) \in \bI : t\le t^\text{tr}\}$ consists of all events up to time $t^\text{tr}$. The remaining events after time $t^\text{tr}$, denoted by the set $\bI^\text{te}=\bI \setminus \bI^\text{tr}$, is reserved for testing.
%$\bG = (\bV, \bE, \bT, \vec{X})$ and training events $\bI=\left\{(i, j, t)_{m}\right\}_{m=1}^{|\bE|}$, where $t < t^{\prime}$, $t^{\prime}$ is the testing time. 
% When training model or node embedding, we have a temporal graph $\bG$ which only contains temporal edges before time $t^{\prime}$. Formally,  $\forall \ (i, j, t)\ \exists \ \bG = (\bV, \bE, \bT), t < t^{\prime}$. 
Given a candidate triple $(i,j,t)$ for some $t > t^\text{tr}$, the objective is to predict whether a link between nodes $i$ and $j$ is formed at the given future time $t$, \ie, if $(i, j, t) \in \bI^\text{te}$. 
Note that our model can perform temporal link prediction between all nodes, including new nodes not seen during training, due to its inductive nature.
Specifically, in testing, we first generate temporal node representations based on the trained model, and feed them to a downstream logistic regression classifier to predict if a candidate triple is positive or negative. The classifier is trained using a 80\%/20\% train/test split on the testing events, and repeated for five different splits. More details are given in Appendix~\ref{app:task}.
%
%For each temporal graph, node representations are learnt on the graph before time $t^{\prime}$, and prediction is about the edges formed at $t^{\prime}$. For instance, on graph cit-HepTh, we train the model only use edges built before $78th$ timestamp, and we predict the edges built at $78th$ timestamp. And we define the edge's test representation as $|\vec{h}_{i}^{t^{\prime}}-\vec{h}_{k}^{t^{\prime}}|$ \cite{lu2019temporal}, for nodes $i$ and $k$. Edges built at time $t^{\prime}$ are positive ones. As for the negative edges(i.e., there is no edge between two nodes), we randomly sample the same number with positive ones. Then, a logistic regression classifier is trained to do the final temporal link prediction, for our model and all baselines. The training ratio of the logistic regression classifier is 80\%. We test each model five times with five different and fixed random seeds. 
%
%\stitle{Secondary task: Temporal node dynamics prediction.}

We further adopt a secondary task of \emph{temporal node dynamics prediction}. While the training and testing events follow the same setup of the main task, 
we aim to predict the number of new neighbors of a node $i$ at a specific future time $t>t^\text{tr}$. 
Similarly, the first step in testing is to generate temporal node representations based on the trained model, which are then fed into a downstream linear regression model. 
To train the regression model, we randomly split the nodes of the testing events into 80\%/20\% train/test split.
%
%For nodes related to testing edges, we randomly split  80\% of them into the training set, and the remain 20\% as testing set. Then, we train a linear regressor to predict $\Delta \hat{N}_i(t^{\prime})$ for the testing set, taking node representation as input.
%\fang{put down the split for training the regression. also, there is no such thing as a logistic regressor. logistic regression is a classification %model, not a regression model.}

%We consider two regression models---one based on the inner product, and the other based on the linear regression, as we futher elaborate in Sect.~\ref{sec:expt:node}.
%After getting the trained model or node embeddings on temporal graph $\bG$ before testing time $t^{\prime}$, that predicting node $i$ will have how many new links at $t^{\prime}$, is the temporal node dynamics prediction. This task is very meaningful in reality. For instance, in a citation graph, this task means predicting a paper will have how many citations in the future. In a e-commerce graph, it means a user will buy how many items or an item will be purchased by how many users, which is very important to the item sellers, even can directly determine a seller will gain or loss. 

\stitle{Settings of TREND.}
For the temporal GNN, we employ two layers with a ReLU activation. The hidden layer dimension is set to 16 on all datasets. The output dimension is set to 32 on CollegeMsg, 16 on cit-HepTh and 128 on Wikipedia and Taobao, based on the size and complexity of the graph.
The transfer function $\textsc{FCL}_{e}$ employs a sigmoid activation, the estimator of node dynamics $\textsc{FCL}_{n}$ uses a ReLU activation, and  $\alpha^{(i,j,t)}, \beta^{(i,j,t)}$ both employ a LeakyReLU. 
The number of negative samples per positive event is set to $Q=1$. For the final loss function in Eq.~\eqref{eq:overall-optimize}, the coefficient of node loss is set to $\eta_1=0.1$ on Taobao and $\eta_1=0.01$ on other datasets, whereas the coefficient of $L_2$ regularizer is set to $\eta_2=0.001$ on CollegeMsg and cit-HepTh, $\eta_2=0.01$ on Taobao, and $\eta_2=1$ on Wikipedia. Note that we will present an analysis on the impact of the hyperparameters $Q,\eta_1$ and $\eta_2$ in Sect.~\ref{sec:expt:paramstudy}. Lastly, we use the Adam optimizer with the learning rate 0.001.

\stitle{Baselines.}
We compare \model\ with a competitive suit of baselines from three categories. 
%A comprehensive set of competitive baselines from two categories are used to compare with our proposed TREND. 
(1) \emph{Static approaches}: DeepWalk \cite{perozzi2014deepwalk}, Node2vec \cite{grover2016node2vec}, VGAE \cite{kipf2016variational}, GAE \cite{kipf2016variational} and GraphSAGE \cite{hamilton2017inductive}. They train a model or node embedding vectors on the static graph formed from the training events, without considering any temporal information. 
(2) \emph{Temporal approaches}: CTDNE \cite{nguyen2018continuous}, EvolveGCN \cite{pareja2020evolvegcn}, GraphSAGE+T \cite{hamilton2017inductive} and TGAT \cite{xu2020inductive}. They train a model or node embedding vectors on the temporal graph formed from the training events. Note that GraphSAGE+T is a temporal extension of GraphSAGE implemented by us, in which the time decay effect is incorporated into the aggregation function. 
(3) \emph{Hawkes process-based approaches}: HTNE \cite{zuo2018embedding} and MMDNE \cite{lu2019temporal}. They similarly train node embedding vectors on the temporal graph formed from the training events. However, they leverage the node representations to model the conditional intensity of events based on the Hawkes process. More baseline descriptions are in Appendix~\ref{app:baselines}.

\subsection{Temporal Link Prediction}
In Tab.~\ref{table:main}, we compare the performance of TREND with the baselines on the main task. In general, our method performs the best among all methods, demonstrating the benefits of event and node dynamics. We make two further observations.

First, among static methods, 
% performances of these baselines are inferior than  those of temporal approaches or Hawkes process approaches, in general. It demonstrates that effectiveness of temporal information and Hawkes process. Among these static approaches,
we can see that GNN-based methods (VGAE, GAE and GraphSAGE)  tend to perform better, as they are inductive in nature, and their message passing scheme is able to  integrate both node features and graph structures. On the other hand, DeepWalk and Node2vec are transductive, which cannot directly extend to new nodes in testing. In our experiments, the embedding vector of new nodes are randomly initialized for tranductive methods, and thus their performance can be poor when dealing with new nodes. One exception is on the CollegeMsg dataset, where there is no node features and one-hot encoding of node IDs are used instead. In this case, GNN-based methods lose the inductive capability and do not outperform transductive methods.
%Because GNN learns aggregation functions to aggregate node features from local structure, being inherently inductive. While network embedding methods, including DeepWalk and Node2vec, only utilize the local structure to directly parameterize the existing node embeddings. They are transductive and cannot deal with new nodes not seen during training. The reason for why network embedding methods perform better on CollegeMsg is that CollegeMsg is a graph having no node features, which will largely undermine the power of GNNs. 

Second, temporal approaches are generally superior to static approaches, showing the importance of temporal information. Among the three GNN-based approaches (EvolveGCN, GraphSAGE+T and TGAT), EvolveGCN often performs the worst. The reason is that EvolveGCN is based on discrete snapshots, which inevitably suffers from some loss in the temporal evolution. 
%Meanwhile, GraphSAGE+T and TGAT model the continuous process of temporal evolution by treating each event as an individual training instance, to capture the temporal dynamics in a more fine-grained manner. 
Moreover, the Hawkes process-based approaches (HTNE and MMDNE) achieve strong performance, demonstrating that the Hawkess process is ideal for modeling the temporal evolution on graphs. Unfortunately, they are transductive and thus do not outperform GraphSAGE-T and TGAT on cit-HepTh and Taobao where there are a large proportion of new nodes in testing. Besides, we can see that TREND performs much better on Taobao than on other datasets. A potential reason is that Taobao is the biggest graph having more ``diversity'' in events, such that the adaptation of event prior becomes more crucial and can lead to larger performance gain.

\subsection{Ablation Study}\label{sec:expt:ablation}
%The strength of our proposed TREND comes from the three components modeling the event and node-dynamics. 
To understand the contribution of each component in \model, we study the following ablated models on the task of temporal link prediction. (1) \emph{TGNN}, which only stacks two temporal GNN layers and optimizes the inner product of node pairs; (2) \emph{TGNN+H}, which adds the global transfer function for the Hawkess process in Eq.~\eqref{eq:transfer function} to \emph{TGNN}; (3) \emph{TGNN+H+E} and \emph{TGNN+H+N}, which further model the event and node dynamics on top of \emph{TGNN+H}, respectively.
Note that \emph{TGNN+H+E} uses the event-specific transfer function in Eq.~\eqref{eq:lambda}.
%it can be generated by adding the event-conditioned transformation in E.q~ \eqref{eq:event transform}, based on \emph{TGNN + H}. (4) \emph{TGNN+H+N}: we can obtain this by adding node dynamics predictor in E.q~ \eqref{eq:node}, based on \emph{TGNN + H}. 
%\fang{renaming the methods:
%1. TGNN (Temporal gnn aggregation)\\
%2. TGNN+H (Temporal gnn + Hawkes process - i.e. the event decode (FCL) as the transfer function )\\
%3. TGNN+H+N (temporal gnn + hawkes + node dynamics)\\
%4. Add another ablation study (TGNN+H+E, temporal gnn + hawkes + event dynamics with conditioning)
%}

\begin{figure}
   \vspace{-2mm}
   \subfigure[Accuracy]{
   \centering
   \includegraphics[width=0.45\linewidth]{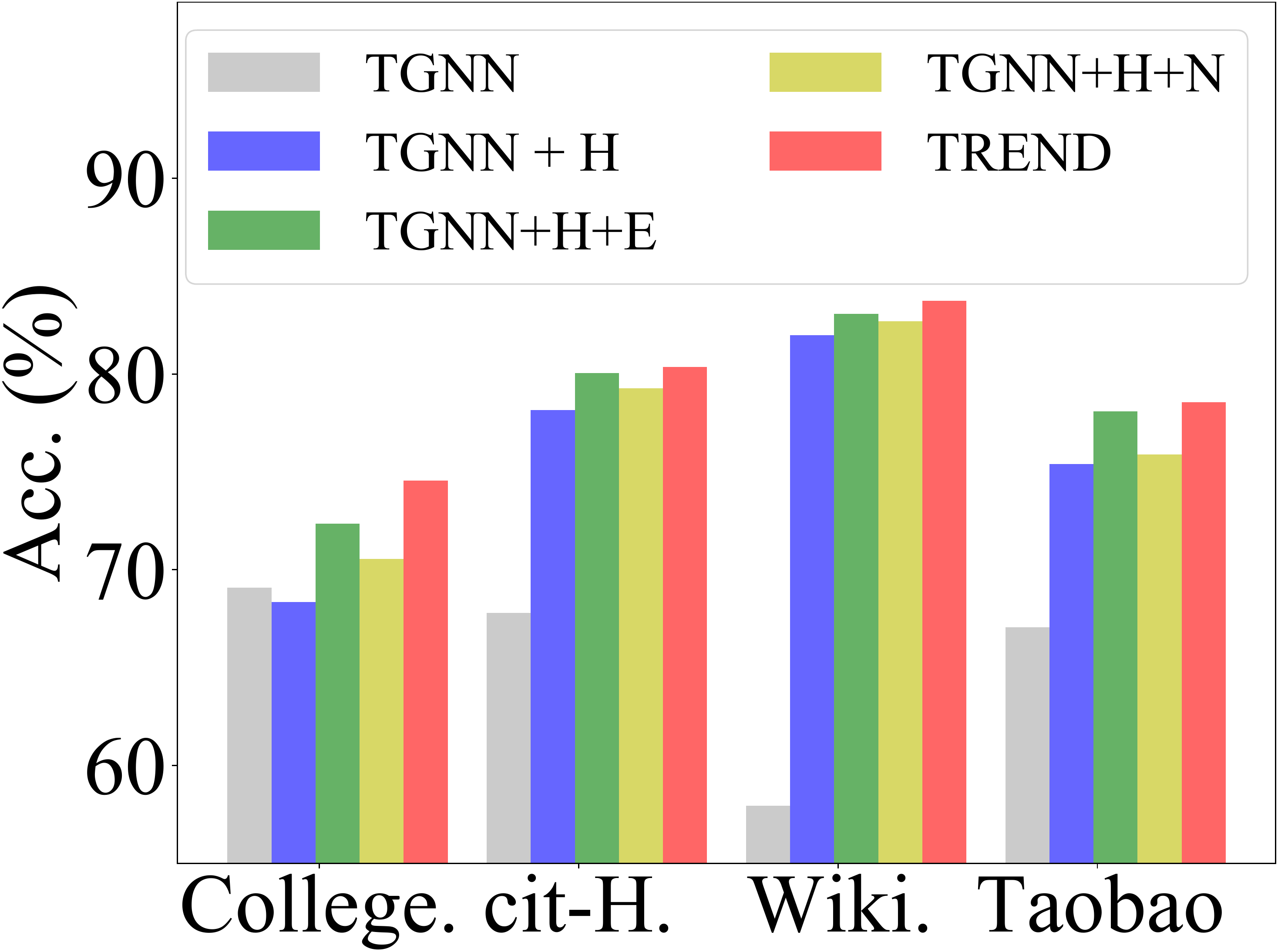}
   }
   \subfigure[F1]{
   \centering
   \includegraphics[width=0.45\linewidth]{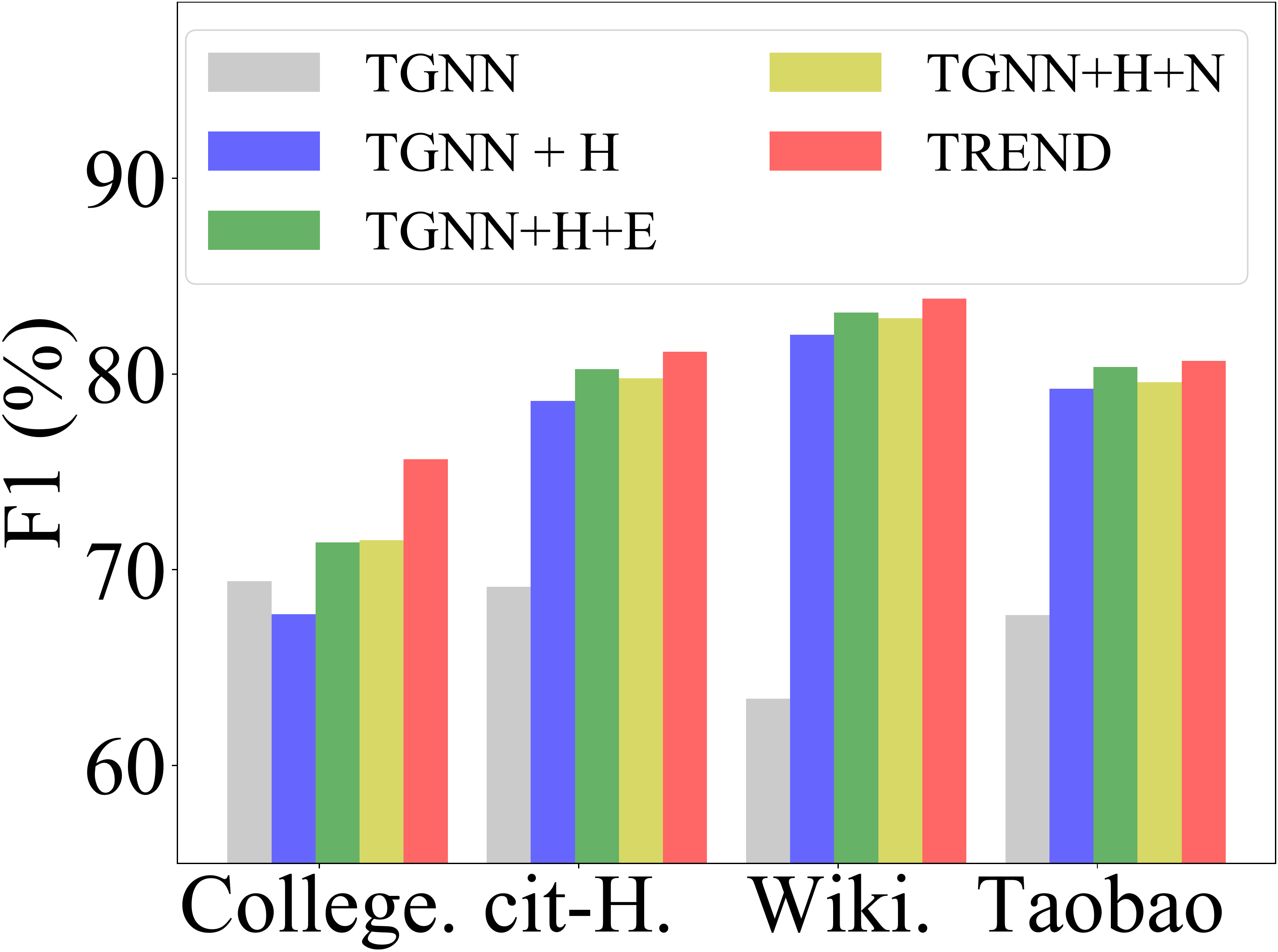}
   }
   \vspace{-4mm}
	\caption{Effect of main components.}
	\vspace{-5mm}
	\label{fig:ablation}
\end{figure}

As shown in Fig.~\ref{fig:ablation}, the performance generally increases when we gradually add more components to \emph{TGNN}. This shows that every component is useful for modeling temporal graphs.
Note that \emph{TGNN+H+E} typically outperforms \emph{TGNN+H+N}, since the event dynamics directly deals with individual events while the node dynamics only works at the node level. Nevertheless, when integrating both event and node dynamics, the full model \model~achieves the best performance, showing that it is important to jointly model both event and node dynamics.
%shows the benefit of our node predictor, \ie, integrating node dynamics will provide helpful evolutionary information to enhance the node representation quality.
%Among the ablated variants, we can notice that there is an obvious improvement in general, when comparing \emph{TGNN + H} and \emph{TGNN}. It means that the transfer function for Hawkes process, works.  The effectiveness of event-conditioned transformation is shown by that \emph{TGNN+H+E} outperforms \emph{TGNN + H}. That \emph{TGNN+H+N} outperforms \emph{TGNN + H} shows the benefit of our node predictor, \ie, integrating node dynamics will provide helpful evolutionary information to enhance the node representation quality.
% TREND outperforms 'With Edge \& Node models' shows that the event-condition prior really works and it is of importance to make customized transfer function for different edges. 
%In general, our TREND outperforms all the ablated variants consistently, demonstrating the overall benefit of the three components modeling the event- and node- dynamics, and that all components are well integrated. 

\subsection{Hyperparameter Study}\label{sec:expt:paramstudy}
Here we present a sensitivity analysis of the hyperparameters. %As for scalability study, it is shown in Appendix~\ref{app:scala}. 
% Here we present a sensitivity analysis of the hyperparameters, and a scalability study on the task of temporal link prediction. 
%study the influence of negative sampling number $Q$ in E.q~\eqref{eq:event loss}, the regularization on the shifting and scaling vectors to restrict the event-conditioned transformation, the constraint $\eta_1$ for node-dynamics, and time complexity.

\begin{figure}
   \subfigure[Negative samples]{
   \centering
   \includegraphics[width=0.305\linewidth]{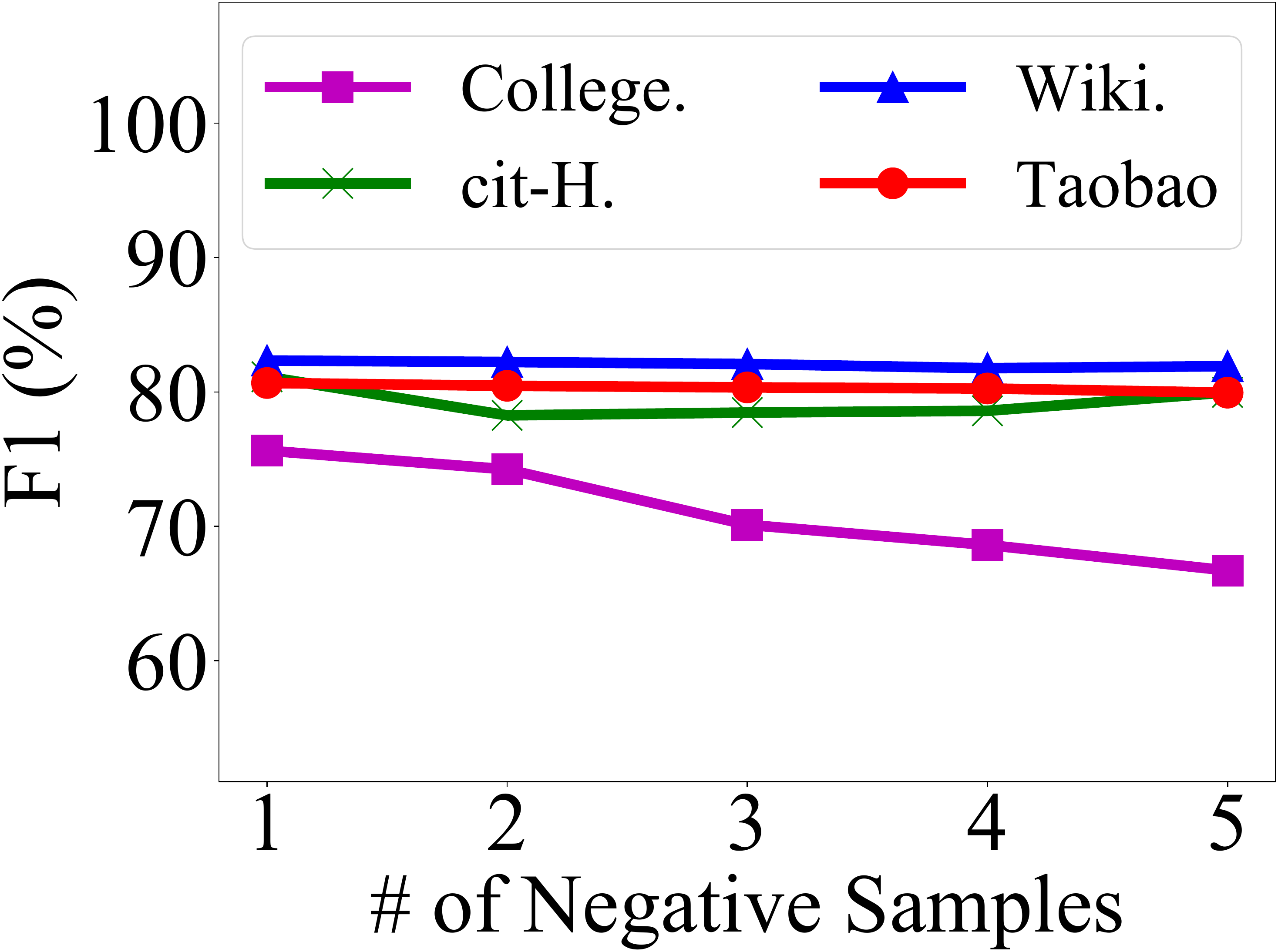}
   }
   \subfigure[Regularization]{
   \centering
   \includegraphics[width=0.305\linewidth]{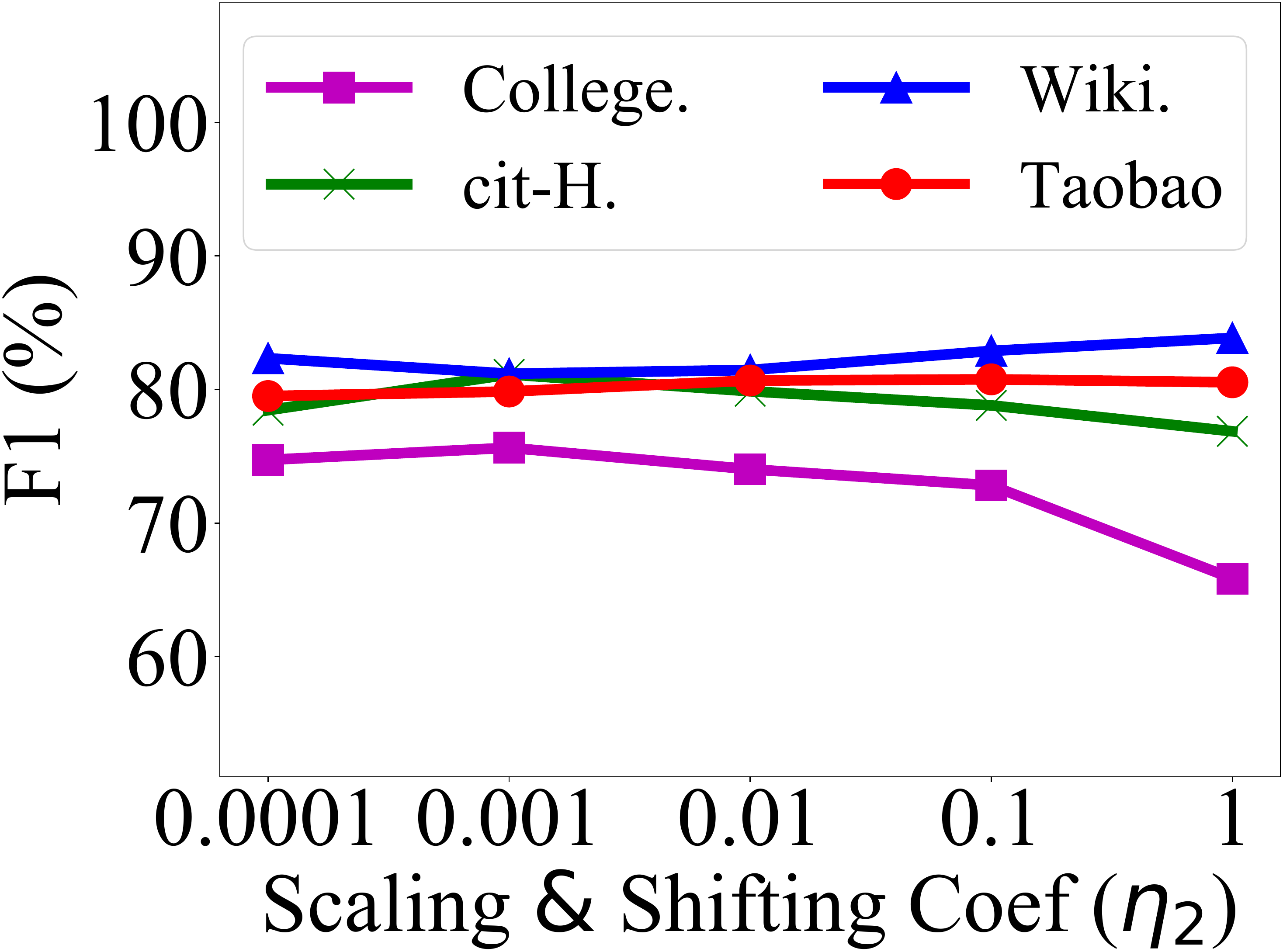}
   }
%   \vspace{-2mm}
   \subfigure[Node coefficient]{
   \centering
   \includegraphics[width=0.305\linewidth]{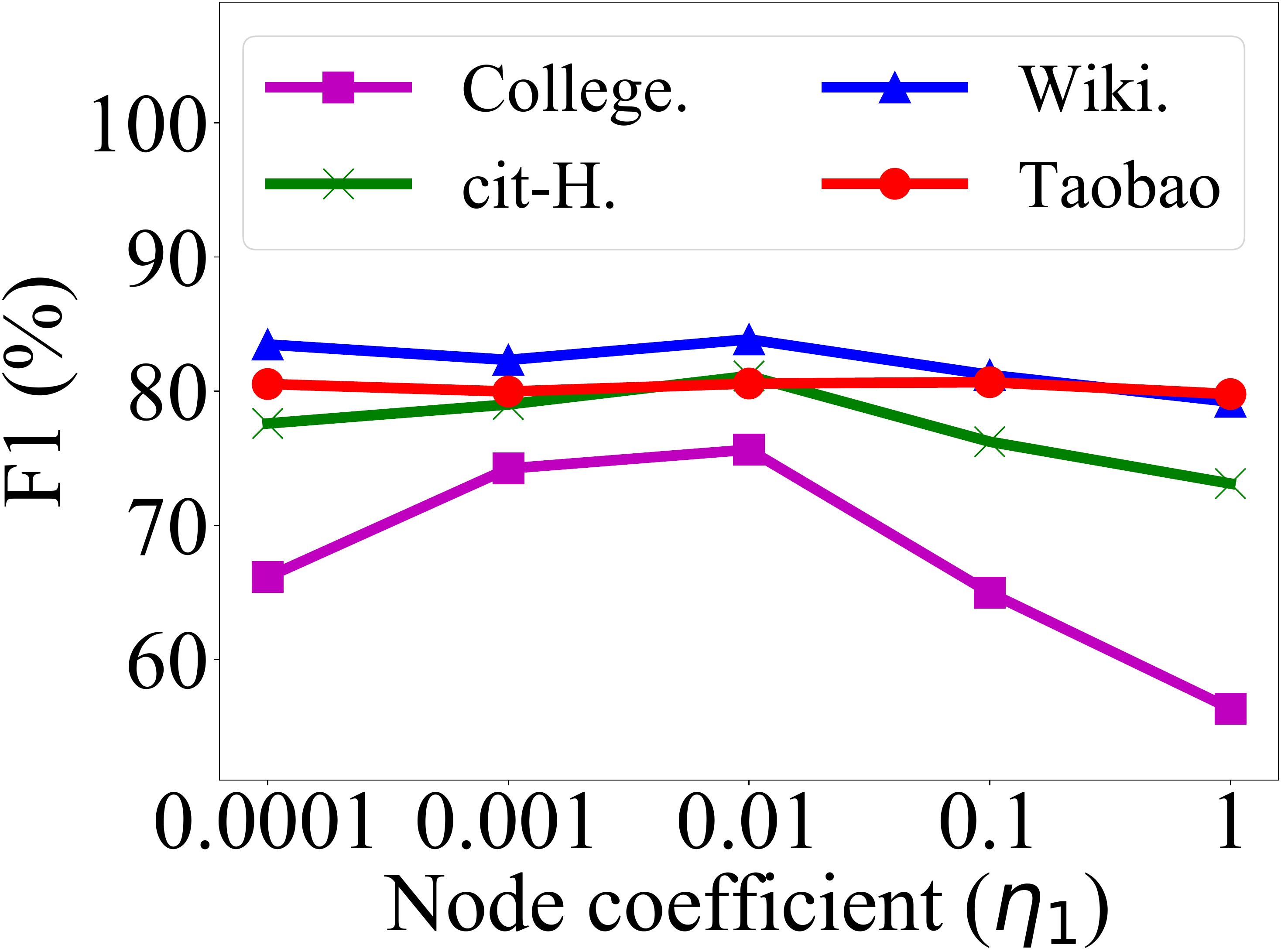}
   }
%   \subfigure[Time complexity]{
%   \centering
%   \includegraphics[width=0.45\linewidth]{figures/time.pdf}
%   }
  \vspace{-4mm}
    \caption{Hyperparameter sensitivity.}
	\label{fig:param}
	\vspace{-2mm}
% 	\caption{Hyperparameter sensitivity, time complexity.}
% 	\label{fig:param}
% 	\vspace{-3mm}
\end{figure}

\stitle{Negative sampling size.}
As shown in Fig.~\ref{fig:param}(a), generally speaking, the performance of \model\ does not improve with more negative samples. 
%Even on CollegeMsg, our model decreases a little. 
On all datasets, it is robust to choose just one negative sample for each positive event for efficiency. 
%s  to In general, our model is not sensitive to the number of negative samples. Therefore, using just 1 negative sample is enough for efficiency reasons. 

\stitle{Regularization on scaling and shifting.}
To prevent overfitting, the event-conditioned transformations are regularized to prevent excessive scaling or shifting. 
%being close to 1 and the shifting being close to 0, to prevent overfitting to the training events. 
The regularization is controlled by the coefficient $\eta_2$, and we study its effect in 
%in E.q~\eqref{eq:overall-optimize}, the effect of regularization is studied in 
Fig.~\ref{fig:param}(b). The performance is quite stable over different values of $\eta_2$, although smaller values in the range [0.0001, 0.01] tend to perform better. Worse performance can be observed on larger values, which implies very little scaling and shifting similar to removing the event-conditioned transformation.% This also reflects the benefit of the even-conditioned transformation.

\stitle{Coefficient for node loss.}
We vary $\eta_1$, which controls the weight of the node loss, and study its impact in Fig.~\ref{fig:param}(c).
We observe that the performance is suboptimal if $\eta_1$ is too small or large, and the performance of \model\ is generally robust when $\eta_1$ is round 0.01. This shows that a well balanced node and event loss can improve the stability and performance. 
%To make the model pay proper attention to capture the node-dynamics, we use $\eta_1$ to constrain $L_{n}$. As shown in Fig.~\ref{fig:param}(c), the performance is not good if $\eta_1$ is too small or too big, and we notice that when $\eta_1$ is round 0.01, TREND can achieve the best performance. The reason are as follows. When $\eta_1$ is too small, the model pay little attention to capture the node-dynamics and cannot utilize the information from node-dynamics. When $\eta_1$ is too big, the whole model pay most attention to node dynamics and ignores the basic event-dynamics.

\begin{table}[t]
   \small
% \centering 
% \renewcommand{\arraystretch}{1.1}
% \vspace{-3mm}
\caption{Performance of node dynamics prediction.
%performance comparison of TREND and baselines, with \emph{MAE} as the metric. Here '-' means that we didn't get node representations.
% \fang{combine table 3 and 4 into one wide table.}
} 
\vspace{-2mm}
\label{table3} 
%\resizebox*{\textwidth}{!}{}{
\begin{tabular}{@{}c|c|c|c|c@{}}  
		\toprule
		 Model&CollegeMsg&cit-HepTh&Wikipedia&Taobao \\\midrule
		CTDNE &10.0636&3.0173&7.3265&0.5789 \\
		EvolveGCN &3.1964&2.5610&6.8651&- \\
		GraphSAGE+T  &21.9444&\underline{2.2421}&\underline{5.9231}&\underline{0.5505} \\
		TGAT &\underline{2.6903}&2.8094&7.7737&0.5550 \\
		HTNE &12.3587&3.2781&6.8860&0.5749 \\
		MMDNE &8.0555&2.7456&6.9552&0.5643 \\
		\midrule
		TREND &\textbf{2.3549}&\textbf{2.2066}&\textbf{5.9140}&\textbf{0.5491}\\
% 		(\% improv.) & (\%) & (\%) & (\%) \\
		\hline
	%\bottomrule
% \vspace{-4mm}
\end{tabular}%}
\vspace{-4mm}
\end{table}
% \vspace{-4mm}

\subsection{Temporal Node Dynamics Prediction}\label{sec:expt:node}
Finally, we evaluate the task of temporal node dynamics prediction. 
%As aforementioned, in order to capture the node-dynamics, we build a node predictor, as shown in E.q~\eqref{eq:node}. In order to know whether it is aware of node-dynamics, we conduct the experiment of temporal-link-number prediction of each node at time $t^{\prime}$, \ie, $\Delta \hat{N}_i(t^{\prime})$. 
% For our TREND, we can directly use our node predictor to compute $\Delta e_{i}^{\prime}(t^{\prime})$. 
% Since all baselines don't  have node predictor, we calculate a probability for each node pairs, defined as $\Phi(i, j) = \sigma(\vec{Z}_{i}^{t^{\prime}} (\vec{Z}_{k}^{t^{\prime}})^{T})$. If $\Phi(i, j) > 0.5$, then we count an edge will connect nodes $i$ and $k$ at future time $t^{\prime}$, following \cite{lu2019temporal}. 
We report the mean absolute error (MAE) between the predicted value $\Delta \hat{N}_i(t)$ and the groundtruth $\Delta N_i(t)$  in Tab.~\ref{table3}. The results show that \model\ consistently achieves the smallest MAE on all four datasets, which demonstrate its versatility beyond temporal link prediction, and that the estimator of node dynamics works well as intended.

\section{Conclusion}
In this paper, we studied the problem of temporal graph representation learning.
Specifically, we proposed \model, a novel framework for temporal graph representation learning, driven by event and node dynamics on a Hawkes process-based GNN. \model\ is inductive and able to capture  a holistic view of the link formation process. More importantly, it integrates both the event and node dynamics to respectively capture the individual and collective characteristics of events, for a more precise modeling of the temporal evolution.
Finally, we conducted extensive experiments on four real-world datasets and demonstrated the superior performance of \model.
%ability of the proposed \model. 
% As for the future direction, maybe we can consider extending our model to multi-graphs. 

% In this paper, we study the problem of temporal link prediction. Specifically, we propose a novel inductive temporal graph representation method (TREND). At event level, with a GNN-based temporal point process, it models the link formation process with customized transfer function, and it constrains the node representation being aware of node-dynamics. And we conduct extensive experiments on three real-world temporal graph datasets, to demonstrate the superior ability on both event- and node-dynamics of the proposed TREND. As for the future direction, maybe we can consider extend our model to pre-training tasks. 

%%
%% The acknowledgments section is defined using the "acks" environment
%% (and NOT an unnumbered section). This ensures the proper
%% identification of the section in the article metadata, and the
%% consistent spelling of the heading.
% \vspace{-2mm}
\begin{acks}
This research is supported by the Agency for Science, Technology and Research (A*STAR) under its AME Programmatic Funds (Grant No. A20H6b0151).
\end{acks}
% \vspace{-4mm}

\clearpage
\balance
%%
%% The next two lines define the bibliography style to be used, and
%% the bibliography file.
\bibliographystyle{ACM-Reference-Format}
\bibliography{references}

%%% -*-BibTeX-*-
%%% Do NOT edit. File created by BibTeX with style
%%% ACM-Reference-Format-Journals [18-Jan-2012].

\begin{thebibliography}{47}

%%% ====================================================================
%%% NOTE TO THE USER: you can override these defaults by providing
%%% customized versions of any of these macros before the \bibliography
%%% command.  Each of them MUST provide its own final punctuation,
%%% except for \shownote{}, \showDOI{}, and \showURL{}.  The latter two
%%% do not use final punctuation, in order to avoid confusing it with
%%% the Web address.
%%%
%%% To suppress output of a particular field, define its macro to expand
%%% to an empty string, or better, \unskip, like this:
%%%
%%% \newcommand{\showDOI}[1]{\unskip}   % LaTeX syntax
%%%
%%% \def \showDOI #1{\unskip}           % plain TeX syntax
%%%
%%% ====================================================================

\ifx \showCODEN    \undefined \def \showCODEN     #1{\unskip}     \fi
\ifx \showDOI      \undefined \def \showDOI       #1{#1}\fi
\ifx \showISBNx    \undefined \def \showISBNx     #1{\unskip}     \fi
\ifx \showISBNxiii \undefined \def \showISBNxiii  #1{\unskip}     \fi
\ifx \showISSN     \undefined \def \showISSN      #1{\unskip}     \fi
\ifx \showLCCN     \undefined \def \showLCCN      #1{\unskip}     \fi
\ifx \shownote     \undefined \def \shownote      #1{#1}          \fi
\ifx \showarticletitle \undefined \def \showarticletitle #1{#1}   \fi
\ifx \showURL      \undefined \def \showURL       {\relax}        \fi
% The following commands are used for tagged output and should be
% invisible to TeX
\providecommand\bibfield[2]{#2}
\providecommand\bibinfo[2]{#2}
\providecommand\natexlab[1]{#1}
\providecommand\showeprint[2][]{arXiv:#2}

\bibitem[\protect\citeauthoryear{Cai, Zheng, and Chang}{Cai
  et~al\mbox{.}}{2018}]%
        {cai2018comprehensive}
\bibfield{author}{\bibinfo{person}{Hongyun Cai}, \bibinfo{person}{Vincent~W
  Zheng}, {and} \bibinfo{person}{Kevin Chen-Chuan Chang}.}
  \bibinfo{year}{2018}\natexlab{}.
\newblock \showarticletitle{A comprehensive survey of graph embedding:
  Problems, techniques, and applications}.
\newblock \bibinfo{journal}{\emph{IEEE Transactions on Knowledge and Data
  Engineering}} \bibinfo{volume}{30}, \bibinfo{number}{9}
  (\bibinfo{year}{2018}), \bibinfo{pages}{1616--1637}.
\newblock


\bibitem[\protect\citeauthoryear{Du, Wang, Song, Lu, and Wang}{Du
  et~al\mbox{.}}{2018}]%
        {du2018dynamic}
\bibfield{author}{\bibinfo{person}{Lun Du}, \bibinfo{person}{Yun Wang},
  \bibinfo{person}{Guojie Song}, \bibinfo{person}{Zhicong Lu}, {and}
  \bibinfo{person}{Junshan Wang}.} \bibinfo{year}{2018}\natexlab{}.
\newblock \showarticletitle{Dynamic network embedding: an extended approach for
  skip-gram based network embedding}. In \bibinfo{booktitle}{\emph{Proceedings
  of the International Joint Conference on Artificial Intelligence}},
  Vol.~\bibinfo{volume}{2018}. \bibinfo{pages}{2086--2092}.
\newblock


\bibitem[\protect\citeauthoryear{Du, Wang, Yang, Zhou, and Tang}{Du
  et~al\mbox{.}}{2019}]%
        {du2019sequential}
\bibfield{author}{\bibinfo{person}{Zhengxiao Du}, \bibinfo{person}{Xiaowei
  Wang}, \bibinfo{person}{Hongxia Yang}, \bibinfo{person}{Jingren Zhou}, {and}
  \bibinfo{person}{Jie Tang}.} \bibinfo{year}{2019}\natexlab{}.
\newblock \showarticletitle{Sequential scenario-specific meta learner for
  online recommendation}. In \bibinfo{booktitle}{\emph{Proceedings of the ACM
  SIGKDD International Conference on Knowledge Discovery and Data Mining}}.
  \bibinfo{pages}{2895--2904}.
\newblock


\bibitem[\protect\citeauthoryear{Girshick}{Girshick}{2015}]%
        {girshick2015fast}
\bibfield{author}{\bibinfo{person}{Ross Girshick}.}
  \bibinfo{year}{2015}\natexlab{}.
\newblock \showarticletitle{Fast {R-CNN}}. In
  \bibinfo{booktitle}{\emph{Proceedings of the IEEE International Conference on
  Computer Vision}}. \bibinfo{pages}{1440--1448}.
\newblock


\bibitem[\protect\citeauthoryear{Goyal, Chhetri, and Canedo}{Goyal
  et~al\mbox{.}}{2020}]%
        {goyal2020dyngraph2vec}
\bibfield{author}{\bibinfo{person}{Palash Goyal}, \bibinfo{person}{Sujit~Rokka
  Chhetri}, {and} \bibinfo{person}{Arquimedes Canedo}.}
  \bibinfo{year}{2020}\natexlab{}.
\newblock \showarticletitle{dyngraph2vec: Capturing network dynamics using
  dynamic graph representation learning}.
\newblock \bibinfo{journal}{\emph{Knowledge-Based Systems}}
  \bibinfo{volume}{187} (\bibinfo{year}{2020}), \bibinfo{pages}{104816}.
\newblock


\bibitem[\protect\citeauthoryear{Goyal, Kamra, He, and Liu}{Goyal
  et~al\mbox{.}}{2018}]%
        {goyal2018dyngem}
\bibfield{author}{\bibinfo{person}{Palash Goyal}, \bibinfo{person}{Nitin
  Kamra}, \bibinfo{person}{Xinran He}, {and} \bibinfo{person}{Yan Liu}.}
  \bibinfo{year}{2018}\natexlab{}.
\newblock \showarticletitle{{DynGEM}: Deep embedding method for dynamic
  graphs}.
\newblock \bibinfo{journal}{\emph{arXiv preprint arXiv:1805.11273}}
  (\bibinfo{year}{2018}).
\newblock


\bibitem[\protect\citeauthoryear{Grover and Leskovec}{Grover and
  Leskovec}{2016}]%
        {grover2016node2vec}
\bibfield{author}{\bibinfo{person}{Aditya Grover} {and} \bibinfo{person}{Jure
  Leskovec}.} \bibinfo{year}{2016}\natexlab{}.
\newblock \showarticletitle{node2vec: Scalable feature learning for networks}.
  In \bibinfo{booktitle}{\emph{Proceedings of the ACM SIGKDD International
  Conference on Knowledge Discovery and Data Mining}}.
  \bibinfo{pages}{855--864}.
\newblock


\bibitem[\protect\citeauthoryear{Ha, Dai, and Le}{Ha et~al\mbox{.}}{2017}]%
        {ha2016hypernetworks}
\bibfield{author}{\bibinfo{person}{David Ha}, \bibinfo{person}{Andrew Dai},
  {and} \bibinfo{person}{Quoc~V Le}.} \bibinfo{year}{2017}\natexlab{}.
\newblock \showarticletitle{Hypernetworks}. In
  \bibinfo{booktitle}{\emph{Proceedings of the International Conference on
  Learning Representations}}.
\newblock


\bibitem[\protect\citeauthoryear{Hajiramezanali, Hasanzadeh, Narayanan,
  Duffield, Zhou, and Qian}{Hajiramezanali et~al\mbox{.}}{2019}]%
        {hajiramezanali2019variational}
\bibfield{author}{\bibinfo{person}{Ehsan Hajiramezanali},
  \bibinfo{person}{Arman Hasanzadeh}, \bibinfo{person}{Krishna Narayanan},
  \bibinfo{person}{Nick Duffield}, \bibinfo{person}{Mingyuan Zhou}, {and}
  \bibinfo{person}{Xiaoning Qian}.} \bibinfo{year}{2019}\natexlab{}.
\newblock \showarticletitle{Variational Graph Recurrent Neural Networks}. In
  \bibinfo{booktitle}{\emph{Proceedings of the International Conference on
  Neural Information Processing Systems}}. \bibinfo{pages}{10701--10711}.
\newblock


\bibitem[\protect\citeauthoryear{Hamilton, Ying, and Leskovec}{Hamilton
  et~al\mbox{.}}{2017}]%
        {hamilton2017inductive}
\bibfield{author}{\bibinfo{person}{William~L Hamilton}, \bibinfo{person}{Rex
  Ying}, {and} \bibinfo{person}{Jure Leskovec}.}
  \bibinfo{year}{2017}\natexlab{}.
\newblock \showarticletitle{Inductive representation learning on large graphs}.
  In \bibinfo{booktitle}{\emph{Proceedings of the International Conference on
  Neural Information Processing Systems}}. \bibinfo{pages}{1025--1035}.
\newblock


\bibitem[\protect\citeauthoryear{Hawkes}{Hawkes}{1971}]%
        {hawkes1971spectra}
\bibfield{author}{\bibinfo{person}{Alan~G Hawkes}.}
  \bibinfo{year}{1971}\natexlab{}.
\newblock \showarticletitle{Spectra of some self-exciting and mutually exciting
  point processes}.
\newblock \bibinfo{journal}{\emph{Biometrika}} \bibinfo{volume}{58},
  \bibinfo{number}{1} (\bibinfo{year}{1971}), \bibinfo{pages}{83--90}.
\newblock


\bibitem[\protect\citeauthoryear{Holme and Saram{\"a}ki}{Holme and
  Saram{\"a}ki}{2012}]%
        {holme2012temporal}
\bibfield{author}{\bibinfo{person}{Petter Holme} {and} \bibinfo{person}{Jari
  Saram{\"a}ki}.} \bibinfo{year}{2012}\natexlab{}.
\newblock \showarticletitle{Temporal networks}.
\newblock \bibinfo{journal}{\emph{Physics reports}} \bibinfo{volume}{519},
  \bibinfo{number}{3} (\bibinfo{year}{2012}), \bibinfo{pages}{97--125}.
\newblock


\bibitem[\protect\citeauthoryear{Ji, Jia, Fang, and Shi}{Ji
  et~al\mbox{.}}{2021}]%
        {DBLP:conf/pkdd/JiJFS21}
\bibfield{author}{\bibinfo{person}{Yugang Ji}, \bibinfo{person}{Tianrui Jia},
  \bibinfo{person}{Yuan Fang}, {and} \bibinfo{person}{Chuan Shi}.}
  \bibinfo{year}{2021}\natexlab{}.
\newblock \showarticletitle{Dynamic Heterogeneous Graph Embedding via
  Heterogeneous Hawkes Process}. In \bibinfo{booktitle}{\emph{Proceedings of
  the European Conference on Machine Learning and Knowledge Discovery in
  Databases, Part {I}}}. \bibinfo{pages}{388--403}.
\newblock


\bibitem[\protect\citeauthoryear{Kingma and Welling}{Kingma and
  Welling}{2014}]%
        {kingma2013auto}
\bibfield{author}{\bibinfo{person}{Diederik~P. Kingma} {and}
  \bibinfo{person}{Max Welling}.} \bibinfo{year}{2014}\natexlab{}.
\newblock \showarticletitle{Auto-Encoding Variational Bayes}. In
  \bibinfo{booktitle}{\emph{Proceedings of the International Conference on
  Learning Representations}}.
\newblock


\bibitem[\protect\citeauthoryear{Kipf and Welling}{Kipf and Welling}{2016}]%
        {kipf2016variational}
\bibfield{author}{\bibinfo{person}{Thomas~N Kipf} {and} \bibinfo{person}{Max
  Welling}.} \bibinfo{year}{2016}\natexlab{}.
\newblock \showarticletitle{Variational graph auto-encoders}.
\newblock \bibinfo{journal}{\emph{arXiv preprint arXiv:1611.07308}}
  (\bibinfo{year}{2016}).
\newblock


\bibitem[\protect\citeauthoryear{Kipf and Welling}{Kipf and Welling}{2017}]%
        {kipf2016semi}
\bibfield{author}{\bibinfo{person}{Thomas~N. Kipf} {and} \bibinfo{person}{Max
  Welling}.} \bibinfo{year}{2017}\natexlab{}.
\newblock \showarticletitle{Semi-Supervised Classification with Graph
  Convolutional Networks}. In \bibinfo{booktitle}{\emph{Proceedings of the
  International Conference on Learning Representations}}.
\newblock


\bibitem[\protect\citeauthoryear{Kumar, Zhang, and Leskovec}{Kumar
  et~al\mbox{.}}{2019}]%
        {kumar2019predicting}
\bibfield{author}{\bibinfo{person}{Srijan Kumar}, \bibinfo{person}{Xikun
  Zhang}, {and} \bibinfo{person}{Jure Leskovec}.}
  \bibinfo{year}{2019}\natexlab{}.
\newblock \showarticletitle{Predicting dynamic embedding trajectory in temporal
  interaction networks}. In \bibinfo{booktitle}{\emph{Proceedings of the ACM
  SIGKDD International Conference on Knowledge Discovery and Data Mining}}.
  \bibinfo{pages}{1269--1278}.
\newblock


\bibitem[\protect\citeauthoryear{Leskovec, Kleinberg, and Faloutsos}{Leskovec
  et~al\mbox{.}}{2005}]%
        {leskovec2005graphs}
\bibfield{author}{\bibinfo{person}{Jure Leskovec}, \bibinfo{person}{Jon
  Kleinberg}, {and} \bibinfo{person}{Christos Faloutsos}.}
  \bibinfo{year}{2005}\natexlab{}.
\newblock \showarticletitle{Graphs over time: densification laws, shrinking
  diameters and possible explanations}. In
  \bibinfo{booktitle}{\emph{Proceedings of the ACM SIGKDD International
  Conference on Knowledge Discovery and Data Mining}}.
  \bibinfo{pages}{177--187}.
\newblock


\bibitem[\protect\citeauthoryear{Li, Cornelius, Liu, Wang, and Barab{\'a}si}{Li
  et~al\mbox{.}}{2017a}]%
        {li2017fundamental}
\bibfield{author}{\bibinfo{person}{Aming Li}, \bibinfo{person}{Sean~P
  Cornelius}, \bibinfo{person}{Y-Y Liu}, \bibinfo{person}{Long Wang}, {and}
  \bibinfo{person}{A-L Barab{\'a}si}.} \bibinfo{year}{2017}\natexlab{a}.
\newblock \showarticletitle{The fundamental advantages of temporal networks}.
\newblock \bibinfo{journal}{\emph{Science}} \bibinfo{volume}{358},
  \bibinfo{number}{6366} (\bibinfo{year}{2017}), \bibinfo{pages}{1042--1046}.
\newblock


\bibitem[\protect\citeauthoryear{Li, Dani, Hu, Tang, Chang, and Liu}{Li
  et~al\mbox{.}}{2017b}]%
        {li2017attributed}
\bibfield{author}{\bibinfo{person}{Jundong Li}, \bibinfo{person}{Harsh Dani},
  \bibinfo{person}{Xia Hu}, \bibinfo{person}{Jiliang Tang}, \bibinfo{person}{Yi
  Chang}, {and} \bibinfo{person}{Huan Liu}.} \bibinfo{year}{2017}\natexlab{b}.
\newblock \showarticletitle{Attributed network embedding for learning in a
  dynamic environment}. In \bibinfo{booktitle}{\emph{Proceedings of the ACM
  Conference on Information and Knowledge Management}}.
  \bibinfo{pages}{387--396}.
\newblock


\bibitem[\protect\citeauthoryear{Li, Zhang, Philip, Zhang, and Yan}{Li
  et~al\mbox{.}}{2018}]%
        {li2018deep}
\bibfield{author}{\bibinfo{person}{Taisong Li}, \bibinfo{person}{Jiawei Zhang},
  \bibinfo{person}{S~Yu Philip}, \bibinfo{person}{Yan Zhang}, {and}
  \bibinfo{person}{Yonghong Yan}.} \bibinfo{year}{2018}\natexlab{}.
\newblock \showarticletitle{Deep dynamic network embedding for link
  prediction}.
\newblock \bibinfo{journal}{\emph{IEEE Access}}  \bibinfo{volume}{6}
  (\bibinfo{year}{2018}), \bibinfo{pages}{29219--29230}.
\newblock


\bibitem[\protect\citeauthoryear{Lu, Wang, Shi, Yu, and Ye}{Lu
  et~al\mbox{.}}{2019}]%
        {lu2019temporal}
\bibfield{author}{\bibinfo{person}{Yuanfu Lu}, \bibinfo{person}{Xiao Wang},
  \bibinfo{person}{Chuan Shi}, \bibinfo{person}{Philip~S Yu}, {and}
  \bibinfo{person}{Yanfang Ye}.} \bibinfo{year}{2019}\natexlab{}.
\newblock \showarticletitle{Temporal network embedding with micro-and
  macro-dynamics}. In \bibinfo{booktitle}{\emph{Proceedings of the ACM
  International Conference on Information and Knowledge Management}}.
  \bibinfo{pages}{469--478}.
\newblock


\bibitem[\protect\citeauthoryear{Mei and Eisner}{Mei and Eisner}{2017}]%
        {mei2017neural}
\bibfield{author}{\bibinfo{person}{Hongyuan Mei} {and} \bibinfo{person}{Jason~M
  Eisner}.} \bibinfo{year}{2017}\natexlab{}.
\newblock \showarticletitle{The Neural Hawkes Process: A Neurally
  Self-Modulating Multivariate Point Process}. In
  \bibinfo{booktitle}{\emph{Proceedings of the ACM SIGKDD International
  Conference on Knowledge Discovery and Data Mining}}.
  \bibinfo{pages}{6754--6764}.
\newblock


\bibitem[\protect\citeauthoryear{Mikolov, Sutskever, Chen, Corrado, and
  Dean}{Mikolov et~al\mbox{.}}{2013}]%
        {mikolov2013distributed}
\bibfield{author}{\bibinfo{person}{Tomas Mikolov}, \bibinfo{person}{Ilya
  Sutskever}, \bibinfo{person}{Kai Chen}, \bibinfo{person}{Greg~S Corrado},
  {and} \bibinfo{person}{Jeff Dean}.} \bibinfo{year}{2013}\natexlab{}.
\newblock \showarticletitle{Distributed representations of words and phrases
  and their compositionality}. In \bibinfo{booktitle}{\emph{Proceedings of the
  International Conference on Neural Information Processing Systems}}.
  \bibinfo{pages}{3111--3119}.
\newblock


\bibitem[\protect\citeauthoryear{Nguyen, Lee, Rossi, Ahmed, Koh, and
  Kim}{Nguyen et~al\mbox{.}}{2018}]%
        {nguyen2018continuous}
\bibfield{author}{\bibinfo{person}{Giang~Hoang Nguyen},
  \bibinfo{person}{John~Boaz Lee}, \bibinfo{person}{Ryan~A Rossi},
  \bibinfo{person}{Nesreen~K Ahmed}, \bibinfo{person}{Eunyee Koh}, {and}
  \bibinfo{person}{Sungchul Kim}.} \bibinfo{year}{2018}\natexlab{}.
\newblock \showarticletitle{Continuous-time dynamic network embeddings}. In
  \bibinfo{booktitle}{\emph{Companion Proceedings of the The Web Conference}}.
  \bibinfo{pages}{969--976}.
\newblock


\bibitem[\protect\citeauthoryear{Panzarasa, Opsahl, and Carley}{Panzarasa
  et~al\mbox{.}}{2009}]%
        {panzarasa2009patterns}
\bibfield{author}{\bibinfo{person}{Pietro Panzarasa}, \bibinfo{person}{Tore
  Opsahl}, {and} \bibinfo{person}{Kathleen~M Carley}.}
  \bibinfo{year}{2009}\natexlab{}.
\newblock \showarticletitle{Patterns and dynamics of users' behavior and
  interaction: Network analysis of an online community}.
\newblock \bibinfo{journal}{\emph{Journal of the American Society for
  Information Science and Technology}} \bibinfo{volume}{60},
  \bibinfo{number}{5} (\bibinfo{year}{2009}), \bibinfo{pages}{911--932}.
\newblock


\bibitem[\protect\citeauthoryear{Pareja, Domeniconi, Chen, Ma, Suzumura,
  Kanezashi, Kaler, Schardl, and Leiserson}{Pareja et~al\mbox{.}}{2020}]%
        {pareja2020evolvegcn}
\bibfield{author}{\bibinfo{person}{Aldo Pareja}, \bibinfo{person}{Giacomo
  Domeniconi}, \bibinfo{person}{Jie Chen}, \bibinfo{person}{Tengfei Ma},
  \bibinfo{person}{Toyotaro Suzumura}, \bibinfo{person}{Hiroki Kanezashi},
  \bibinfo{person}{Tim Kaler}, \bibinfo{person}{Tao Schardl}, {and}
  \bibinfo{person}{Charles Leiserson}.} \bibinfo{year}{2020}\natexlab{}.
\newblock \showarticletitle{Evolvegcn: Evolving graph convolutional networks
  for dynamic graphs}. In \bibinfo{booktitle}{\emph{Proceedings of the AAAI
  Conference on Artificial Intelligence}}. \bibinfo{pages}{5363--5370}.
\newblock


\bibitem[\protect\citeauthoryear{Pennebaker, Francis, and Booth}{Pennebaker
  et~al\mbox{.}}{2001}]%
        {pennebaker2001linguistic}
\bibfield{author}{\bibinfo{person}{James~W Pennebaker},
  \bibinfo{person}{Martha~E Francis}, {and} \bibinfo{person}{Roger~J Booth}.}
  \bibinfo{year}{2001}\natexlab{}.
\newblock \bibinfo{booktitle}{\emph{Linguistic inquiry and word count: LIWC
  2001}}.
\newblock \bibinfo{publisher}{Mahway: Lawrence Erlbaum Associates}.
\newblock


\bibitem[\protect\citeauthoryear{Perez, Strub, De~Vries, Dumoulin, and
  Courville}{Perez et~al\mbox{.}}{2018}]%
        {perez2018film}
\bibfield{author}{\bibinfo{person}{Ethan Perez}, \bibinfo{person}{Florian
  Strub}, \bibinfo{person}{Harm De~Vries}, \bibinfo{person}{Vincent Dumoulin},
  {and} \bibinfo{person}{Aaron Courville}.} \bibinfo{year}{2018}\natexlab{}.
\newblock \showarticletitle{{FiLM}: Visual reasoning with a general
  conditioning layer}. In \bibinfo{booktitle}{\emph{Proceedings of the AAAI
  Conference on Artificial Intelligence}}.
\newblock


\bibitem[\protect\citeauthoryear{Perozzi, Al-Rfou, and Skiena}{Perozzi
  et~al\mbox{.}}{2014}]%
        {perozzi2014deepwalk}
\bibfield{author}{\bibinfo{person}{Bryan Perozzi}, \bibinfo{person}{Rami
  Al-Rfou}, {and} \bibinfo{person}{Steven Skiena}.}
  \bibinfo{year}{2014}\natexlab{}.
\newblock \showarticletitle{{DeepWalk}: Online learning of social
  representations}. In \bibinfo{booktitle}{\emph{Proceedings of the ACM SIGKDD
  international conference on Knowledge discovery and data mining}}.
  \bibinfo{pages}{701--710}.
\newblock


\bibitem[\protect\citeauthoryear{Rezende, Mohamed, and Wierstra}{Rezende
  et~al\mbox{.}}{2014}]%
        {rezende2014stochastic}
\bibfield{author}{\bibinfo{person}{Danilo~Jimenez Rezende},
  \bibinfo{person}{Shakir Mohamed}, {and} \bibinfo{person}{Daan Wierstra}.}
  \bibinfo{year}{2014}\natexlab{}.
\newblock \showarticletitle{Stochastic backpropagation and approximate
  inference in deep generative models}. In
  \bibinfo{booktitle}{\emph{Proceedings of the International Conference on
  Machine Learning}}. \bibinfo{pages}{1278--1286}.
\newblock


\bibitem[\protect\citeauthoryear{Sankar, Wu, Gou, Zhang, and Yang}{Sankar
  et~al\mbox{.}}{2020}]%
        {sankar2020dysat}
\bibfield{author}{\bibinfo{person}{Aravind Sankar}, \bibinfo{person}{Yanhong
  Wu}, \bibinfo{person}{Liang Gou}, \bibinfo{person}{Wei Zhang}, {and}
  \bibinfo{person}{Hao Yang}.} \bibinfo{year}{2020}\natexlab{}.
\newblock \showarticletitle{{DySAT}: Deep neural representation learning on
  dynamic graphs via self-attention networks}. In
  \bibinfo{booktitle}{\emph{Proceedings of the ACM International Conference on
  Web Search and Data Mining}}. \bibinfo{pages}{519--527}.
\newblock


\bibitem[\protect\citeauthoryear{Sarkar, Chakrabarti, and Jordan}{Sarkar
  et~al\mbox{.}}{2012}]%
        {sarkar2012nonparametric}
\bibfield{author}{\bibinfo{person}{Purnamrita Sarkar},
  \bibinfo{person}{Deepayan Chakrabarti}, {and} \bibinfo{person}{Michael~I
  Jordan}.} \bibinfo{year}{2012}\natexlab{}.
\newblock \showarticletitle{Nonparametric Link Prediction in Dynamic Networks}.
  In \bibinfo{booktitle}{\emph{Proceedings of the International Conference on
  Machine Learning}}. \bibinfo{pages}{1897--1904}.
\newblock


\bibitem[\protect\citeauthoryear{Singer, Guy, and Radinsky}{Singer
  et~al\mbox{.}}{2019}]%
        {ijcai2019-640}
\bibfield{author}{\bibinfo{person}{Uriel Singer}, \bibinfo{person}{Ido Guy},
  {and} \bibinfo{person}{Kira Radinsky}.} \bibinfo{year}{2019}\natexlab{}.
\newblock \showarticletitle{Node Embedding over Temporal Graphs}. In
  \bibinfo{booktitle}{\emph{Proceedings of the International Joint Conference
  on Artificial Intelligence}}. \bibinfo{pages}{4605--4612}.
\newblock


\bibitem[\protect\citeauthoryear{Tang, Qu, Wang, Zhang, Yan, and Mei}{Tang
  et~al\mbox{.}}{2015}]%
        {tang2015line}
\bibfield{author}{\bibinfo{person}{Jian Tang}, \bibinfo{person}{Meng Qu},
  \bibinfo{person}{Mingzhe Wang}, \bibinfo{person}{Ming Zhang},
  \bibinfo{person}{Jun Yan}, {and} \bibinfo{person}{Qiaozhu Mei}.}
  \bibinfo{year}{2015}\natexlab{}.
\newblock \showarticletitle{{LINE}: Large-scale information network embedding}.
  In \bibinfo{booktitle}{\emph{Proceedings of the International Conference on
  World Wide Web}}. \bibinfo{pages}{1067--1077}.
\newblock


\bibitem[\protect\citeauthoryear{Trivedi, Farajtabar, Biswal, and Zha}{Trivedi
  et~al\mbox{.}}{2019}]%
        {trivedi2019dyrep}
\bibfield{author}{\bibinfo{person}{Rakshit Trivedi}, \bibinfo{person}{Mehrdad
  Farajtabar}, \bibinfo{person}{Prasenjeet Biswal}, {and}
  \bibinfo{person}{Hongyuan Zha}.} \bibinfo{year}{2019}\natexlab{}.
\newblock \showarticletitle{{DyRep}: Learning representations over dynamic
  graphs}. In \bibinfo{booktitle}{\emph{Proceedings of the International
  Conference on Learning Representations}}.
\newblock


\bibitem[\protect\citeauthoryear{Veličković, Cucurull, Casanova, Romero,
  Liò, and Bengio}{Veličković et~al\mbox{.}}{2018}]%
        {velic2018graph}
\bibfield{author}{\bibinfo{person}{Petar Veličković},
  \bibinfo{person}{Guillem Cucurull}, \bibinfo{person}{Arantxa Casanova},
  \bibinfo{person}{Adriana Romero}, \bibinfo{person}{Pietro Liò}, {and}
  \bibinfo{person}{Yoshua Bengio}.} \bibinfo{year}{2018}\natexlab{}.
\newblock \showarticletitle{Graph Attention Networks}. In
  \bibinfo{booktitle}{\emph{Proceedings of the International Conference on
  Learning Representations}}.
\newblock


\bibitem[\protect\citeauthoryear{Wang, Chang, Liu, Leskovec, and Li}{Wang
  et~al\mbox{.}}{2021}]%
        {wang2021inductive}
\bibfield{author}{\bibinfo{person}{Yanbang Wang}, \bibinfo{person}{Yen-Yu
  Chang}, \bibinfo{person}{Yunyu Liu}, \bibinfo{person}{Jure Leskovec}, {and}
  \bibinfo{person}{Pan Li}.} \bibinfo{year}{2021}\natexlab{}.
\newblock \showarticletitle{Inductive Representation Learning in Temporal
  Networks via Causal Anonymous Walks}. In
  \bibinfo{booktitle}{\emph{Proceedings of the International Conference on
  Learning Representations}}.
\newblock


\bibitem[\protect\citeauthoryear{Wen, Fang, and Liu}{Wen et~al\mbox{.}}{2021}]%
        {wen2021meta}
\bibfield{author}{\bibinfo{person}{Zhihao Wen}, \bibinfo{person}{Yuan Fang},
  {and} \bibinfo{person}{Zemin Liu}.} \bibinfo{year}{2021}\natexlab{}.
\newblock \showarticletitle{Meta-Inductive Node Classification across Graphs}.
  In \bibinfo{booktitle}{\emph{Proceedings of the International {ACM} {SIGIR}
  Conference on Research and Development in Information Retrieval}}.
  \bibinfo{pages}{1219--1228}.
\newblock


\bibitem[\protect\citeauthoryear{Wu, Souza, Zhang, Fifty, Yu, and
  Weinberger}{Wu et~al\mbox{.}}{2019}]%
        {wu2019simplifying}
\bibfield{author}{\bibinfo{person}{Felix Wu}, \bibinfo{person}{Amauri Souza},
  \bibinfo{person}{Tianyi Zhang}, \bibinfo{person}{Christopher Fifty},
  \bibinfo{person}{Tao Yu}, {and} \bibinfo{person}{Kilian Weinberger}.}
  \bibinfo{year}{2019}\natexlab{}.
\newblock \showarticletitle{Simplifying graph convolutional networks}. In
  \bibinfo{booktitle}{\emph{Proceedings of the International Conference on
  Machine Learning}}. \bibinfo{pages}{6861--6871}.
\newblock


\bibitem[\protect\citeauthoryear{Wu, Pan, Chen, Long, Zhang, and Philip}{Wu
  et~al\mbox{.}}{2020}]%
        {wu2020comprehensive}
\bibfield{author}{\bibinfo{person}{Zonghan Wu}, \bibinfo{person}{Shirui Pan},
  \bibinfo{person}{Fengwen Chen}, \bibinfo{person}{Guodong Long},
  \bibinfo{person}{Chengqi Zhang}, {and} \bibinfo{person}{S~Yu Philip}.}
  \bibinfo{year}{2020}\natexlab{}.
\newblock \showarticletitle{A comprehensive survey on graph neural networks}.
\newblock \bibinfo{journal}{\emph{IEEE Transactions on Neural Networks and
  Learning Systems}} \bibinfo{volume}{32}, \bibinfo{number}{1}
  (\bibinfo{year}{2020}), \bibinfo{pages}{4--24}.
\newblock


\bibitem[\protect\citeauthoryear{Xu, Ruan, K{\"{o}}rpeoglu, Kumar, and
  Achan}{Xu et~al\mbox{.}}{2020}]%
        {xu2020inductive}
\bibfield{author}{\bibinfo{person}{Da Xu}, \bibinfo{person}{Chuanwei Ruan},
  \bibinfo{person}{Evren K{\"{o}}rpeoglu}, \bibinfo{person}{Sushant Kumar},
  {and} \bibinfo{person}{Kannan Achan}.} \bibinfo{year}{2020}\natexlab{}.
\newblock \showarticletitle{Inductive representation learning on temporal
  graphs}. In \bibinfo{booktitle}{\emph{Proceedings of the International
  Conference on Learning Representations}}.
\newblock


\bibitem[\protect\citeauthoryear{Xu, Ba, Kiros, Cho, Courville, Salakhudinov,
  Zemel, and Bengio}{Xu et~al\mbox{.}}{2015}]%
        {xu2015show}
\bibfield{author}{\bibinfo{person}{Kelvin Xu}, \bibinfo{person}{Jimmy Ba},
  \bibinfo{person}{Ryan Kiros}, \bibinfo{person}{Kyunghyun Cho},
  \bibinfo{person}{Aaron Courville}, \bibinfo{person}{Ruslan Salakhudinov},
  \bibinfo{person}{Rich Zemel}, {and} \bibinfo{person}{Yoshua Bengio}.}
  \bibinfo{year}{2015}\natexlab{}.
\newblock \showarticletitle{Show, attend and tell: Neural image caption
  generation with visual attention}. In \bibinfo{booktitle}{\emph{Proceedings
  of the International Conference on Machine learning}}. PMLR,
  \bibinfo{pages}{2048--2057}.
\newblock


\bibitem[\protect\citeauthoryear{Xu, Hu, Leskovec, and Jegelka}{Xu
  et~al\mbox{.}}{2019}]%
        {xu2018powerful}
\bibfield{author}{\bibinfo{person}{Keyulu Xu}, \bibinfo{person}{Weihua Hu},
  \bibinfo{person}{Jure Leskovec}, {and} \bibinfo{person}{Stefanie Jegelka}.}
  \bibinfo{year}{2019}\natexlab{}.
\newblock \showarticletitle{How Powerful are Graph Neural Networks?}. In
  \bibinfo{booktitle}{\emph{Proceedings of the International Conference on
  Learning Representations}}.
\newblock


\bibitem[\protect\citeauthoryear{Zhou, Yang, Ren, Wu, and Zhuang}{Zhou
  et~al\mbox{.}}{2018}]%
        {zhou2018dynamic}
\bibfield{author}{\bibinfo{person}{Lekui Zhou}, \bibinfo{person}{Yang Yang},
  \bibinfo{person}{Xiang Ren}, \bibinfo{person}{Fei Wu}, {and}
  \bibinfo{person}{Yueting Zhuang}.} \bibinfo{year}{2018}\natexlab{}.
\newblock \showarticletitle{Dynamic network embedding by modeling triadic
  closure process}. In \bibinfo{booktitle}{\emph{Proceedings of the AAAI
  Conference on Artificial Intelligence}}.
\newblock


\bibitem[\protect\citeauthoryear{Zhu, Cui, Zhang, Pei, and Zhu}{Zhu
  et~al\mbox{.}}{2018}]%
        {zhu2018high}
\bibfield{author}{\bibinfo{person}{Dingyuan Zhu}, \bibinfo{person}{Peng Cui},
  \bibinfo{person}{Ziwei Zhang}, \bibinfo{person}{Jian Pei}, {and}
  \bibinfo{person}{Wenwu Zhu}.} \bibinfo{year}{2018}\natexlab{}.
\newblock \showarticletitle{High-order proximity preserved embedding for
  dynamic networks}.
\newblock \bibinfo{journal}{\emph{IEEE Transactions on Knowledge and Data
  Engineering}} \bibinfo{volume}{30}, \bibinfo{number}{11}
  (\bibinfo{year}{2018}), \bibinfo{pages}{2134--2144}.
\newblock


\bibitem[\protect\citeauthoryear{Zuo, Liu, Lin, Guo, Hu, and Wu}{Zuo
  et~al\mbox{.}}{2018}]%
        {zuo2018embedding}
\bibfield{author}{\bibinfo{person}{Yuan Zuo}, \bibinfo{person}{Guannan Liu},
  \bibinfo{person}{Hao Lin}, \bibinfo{person}{Jia Guo},
  \bibinfo{person}{Xiaoqian Hu}, {and} \bibinfo{person}{Junjie Wu}.}
  \bibinfo{year}{2018}\natexlab{}.
\newblock \showarticletitle{Embedding temporal network via neighborhood
  formation}. In \bibinfo{booktitle}{\emph{Proceedings of the ACM SIGKDD
  International Conference on Knowledge Discovery and Data Mining}}.
  \bibinfo{pages}{2857--2866}.
\newblock


\end{thebibliography}

\clearpage
\balance

\appendix
\section*{Appendices}

\section{Connection between transfer function and conditional intensity }\label{app:proof}
In the following, we show that a well chosen transfer function $f$, taking the temporal representations as input, is equivalent to the conditional intensity of the Hawkes process in Eq.~\eqref{eq:lambda 0}. Suppose the temporal representations are generated through stacking $l$ temporal GNN layers. First, let us define the base intensity as a function of the self-information:
\begin{align}
    \mu_{i,j}(t)&=f_\mu(\vec{h}_{i}^{t,l-1}\vec{W}_\text{self}^{l}, \vec{h}_{j}^{t,l-1}\vec{W}_\text{self}^{l}).
\end{align}

Next, we define the amount of excitement induced by a historical neighbor as a function of the historical neighbors' information:
\begin{align}
    \gamma_{j'}(t')&=f_\gamma(\vec{h}_{j'}^{t', l-1}\vec{W}_\text{hist}^{l}), \quad
    \gamma_{i'}(t')&=f_\gamma(\vec{h}_{i'}^{t', l-1}\vec{W}_\text{hist}^{l}).
\end{align}

Given these building blocks, we rewrite the conditional intensity in Eq.~\eqref{eq:lambda 0} as 
\begin{align}
    \lambda_{i,j}(t)=f_\lambda\Big(&\vec{h}_{i}^{t,l-1}\vec{W}_\text{self}^{l}, \{\vec{h}_{j'}^{t', l-1}\vec{W}_\text{hist}^{l}:(i,j',t')\in \bH_i(t)\},\nonumber\\
    &\vec{h}_{j}^{t,l-1}\vec{W}_\text{self}^{l},\{\vec{h}_{i'}^{t', l-1}\vec{W}_\text{hist}^{l}:(i',j,t')\in \bH_j(t)\}\Big),
\end{align}
where $f_\lambda$ is a composite function of $f_\mu$, $f_\gamma$ and the summation. By choosing the right transfer function $f$, we further rewrite $f_\lambda$ as the composition of $f$ and the temporal GNN layer $f_g$ given in Eq.~\eqref{eq:tgnn}, \ie, $f_\lambda=f\circ f_g$.
Subsequently, the conditional intensity is given by
\begin{align}
    \lambda_{i,j}(t)&=(f\circ f_g)\Big(\vec{h}_{i}^{t,l-1}\vec{W}_\text{self}^{l}, \{\vec{h}_{j'}^{t', l-1}\vec{W}_\text{hist}^{l}:(i,j',t')\in \bH_i(t)\},\nonumber\\
    &\hspace{1.5cm} \vec{h}_{j}^{t,l-1}\vec{W}_\text{self}^{l},\{\vec{h}_{i'}^{t', l-1}\vec{W}_\text{hist}^{l}:(i',j,t')\in \bH_j(t)\}\Big)\nonumber\\
    &=f(\vec{h}_i^t,\vec{h}_j^t).\label{eq:lambda_equivalence}
\end{align}
Thus, a well-fit transfer function $f$, such as a neural network, can approximate the conditional intensity.

\section{Pseudocode}\label{app:pseudocode}

We outline the training procedure of \model\ in Algorithm~\ref{alg:train}.

\begin{algorithm}[h]
\algrenewcommand\algorithmicrequire{\textbf{Input:}}
\algrenewcommand\algorithmicensure{\textbf{Output:}}
\small
\caption{\textsc{Training Procedure of \model}}
\label{alg:train}
\begin{algorithmic}[1]
    \Require Training graph $\bG = (\bV, \bE, \bT, \vec{X})$, training events $\bI^\text{tr}$. 
%    =\left\{(i, j, t)_{m}\right\}_{m=1}^{|\bE|}$, where $t < t^{\prime}$, $t^{\prime}$ is the testing time. 
%; for each graph $G_i$, the corresponding meta-edge vector $\vec{g}_i$. % and their corresponding meta-edge vectors: $ g^\text{tr}$; 
    %testing graphs: $ \bG^\text{te}$ and their corresponding meta-edge vectors: $ g^\text{te}$; 
    %inner task level adaptation learning rate:$ \alpha$; outer  framework learning rate $ \alpha_{out}$.
    % $\forall \ (i, j, t)\ \exists \ $
    \Ensure Temporal GNN $\theta_{g}$, event prior $\theta_{e}$, transformation model $\theta_\tau$, estimator of node dynamics $\theta_{n}$.
    % \Ensure Prior $\Theta$.
    \State $\theta_{g},\theta_{e},\theta_\tau, \theta_{n}\gets$ parameters initialization;
    \While{not converged}
        %\State sample a batch of $\bG_{b}^\text{tr}\subset\bG^\text{tr}$ together with their corresponding meta-edge vectors: $ g_{b}^\text{tr} \subset g^\text{tr}$;
        \State sample a batch of temporal events $(i, j, t)$ from $\bI^\text{tr}$;
        \For{each event $(i, j, t)$ in the batch}
            % \State sample support set $S_i$, query set $Q_i$ from $G_i$;
            \State calculate node representations $\vec{h}_{i}^{t}$, $\vec{h}_{j}^{t}$for nodes $i$,$j$;
            \Comment{Eq.~\eqref{eq:tgnn}}
            % \For{each layer $l\in\{1,...,L\}$}
            \State $\theta_{e}^{(i, j, t)} \gets$ event-conditioned adaptation on $\theta_{e}$; \Comment{Eq.~\eqref{eq:event transform}}
            \State calculate event intensity $\lambda_{i, j}(t)$;          
            \Comment{Eq.~\eqref{eq:lambda}}
            \State calculate %event loss $L_{e}$, node level loss $L_{n}$ , then overall loss; 
            the overall loss;
            \Comment{Eqs.~\eqref{eq:event loss}, ~\eqref{eq:node loss}, ~\eqref{eq:overall-optimize}}
            % \State calculate task (query) loss $L(Q_i,\theta'_{i})$;            %\Comment{Eq.~\eqref{eq:overall-optimize}}
            % \State
            % \Comment
            % \State $\vec{r}^l_{v'} \leftarrow (\gamma^l_{v'} + 1.0)\odot \vec{r}^l + \beta^l_{v'}$; \Comment{Eq.~\ref{eq.r-film}}
            % \EndFor
        \EndFor 
        \State $\theta_{g},\theta_{e},\theta_\tau, \theta_{n}\gets$ backpropagation of overall loss  \Comment{Eq.~\eqref{eq:overall-optimize}}
    \EndWhile 
    % \State 
    % \State \Return $\Theta$.
    \State \Return $\theta_{g},\theta_{e},\theta_\tau, \theta_{n}$.
\end{algorithmic}
\end{algorithm}

\section{Additional Description of Datasets}\label{app:dataset}
We include more details of the datasets below.
\begin{itemize}[leftmargin=*]
    \item CollegeMsg \cite{panzarasa2009patterns} is an online social network where private messages were sent and received at the University of California, Irvine. If user $i$ sent a private message to user $j$ at time $t$, there is a temporal edge $(i, j, t)$. Since the nodes have no feature, we use the one-hot encoding of the node ID as node features.
    \item cit-HepTh \cite{leskovec2005graphs} is  a citation graph about high energy physics theory from the e-print arXiv, in the period from January 1993 to April 2003. A temporal edge $(i, j, t)$ here means a paper $i$ cites paper $j$ at time $t$. We use word2vec \cite{mikolov2013distributed} to convert the text of paper abstract (\ie, the raw node features) into node embedding as the node feature.
    \item Wikipedia \cite{kumar2019predicting} is a graph in which temporal edges are interactions induced by users' editing on the Wikipedia pages in one month. User edits consist of textual features, which are converted into 172-dimensional LIWC \cite{pennebaker2001linguistic} feature vectors. The edit vectors of each user are added and normalized to serve as the node feature.
    \item Taobao \cite{du2019sequential} is a quite large online purchase network on the e-commerce platform taobao.com. If a user $i$ purchased an item $j$ at time $t$, there is  a temporal edge $(i, j, t)$. Node features are preprocessed embeddings of textual features.
\end{itemize}

\section{Details of Task Setup}\label{app:task}

We describe more details of our main task, namely, temporal link prediction.
For each temporal graph, node representations are learnt on the graph consisting of events before time $t^\text{tr}$, and we try to predict events on or after $t^\text{tr}$. In our experiments, we use all the events before the last time step for training,  and test on the events at the last time step.
For instance, on the graph cit-HepTh, we train the model only using events before the 78$^\text{th}$ time step, and we predict the links formed at the 78$^\text{th}$ time step.
At test time, a logistic regression classifier is trained for the downstream task of  temporal link prediction. While links formed at the last time step are our positive examples, we further randomly sample an equal number of  negative examples (\ie, node pairs which do not form a link at the last time step). We define the feature vector of a candidate triple $(i,j,t)$ as 
%we define the edge's test representation as
$|\vec{h}_{i}^{t}-\vec{h}_{j}^{t}|$ \cite{lu2019temporal}. 
%Edges built at time $t^{\prime}$ are positive ones. As for the negative edges(i.e., there is no edge between two nodes), we randomly sample the same number with positive ones. Then, a logistic regression classifier is trained to do the final temporal link prediction, for our model and all baselines. The training ratio of the logistic regression classifier is 80\%. We test each model five times with five different and fixed random seeds. 

%\stitle{Temporal node dynamics prediction.}
%Having trained model or node embedding on temporal graph $\bG$ before time $t$, predicting that node $i$ will have how many new links at $t$, is the temporal link number prediction of node. This task is very meaningful in reality. For instance, in a citation graph, this task means predicting a paper will have how many citations in the future. In a e-commerce graph, it means a user will buy how many items or an item will be purchased by how many users, which is very important to the item sellers, even can directly determine a seller whether will gain or loss. 

%For nodes related to testing edges, we randomly split  80\% of them into the training set, and the remain 20\% as testing set. Then, we train a linear regressor to predict $\Delta \hat{N}_i(t^{\prime})$ for the testing set, taking node representation as input.

\section{Additional Description of Baselines}\label{app:baselines}
We include more details of the baselines below.

(1) \emph{Static approaches}, in which models or node embedding vectors are trained on the static graph formed before the testing time, regardless of the temporal information.
\begin{itemize}[leftmargin=*]
    \item DeepWalk \cite{perozzi2014deepwalk}: a static network embedding method, which regards the random walk sequences as sentences and leverages skip-grams \cite{mikolov2013distributed} to learn node embeddings.
    \item Node2vec \cite{grover2016node2vec}: another static network embedding method, which generalizes DeepWalk with biased random walks.
    \item VGAE \cite{kipf2016variational}: based on variational auto-encoder (VAE) \cite{kingma2013auto, rezende2014stochastic}, it is a classical GNN-based link prediction model, using a GCN \cite{kipf2016semi} encoder and an inner product decoder.
    \item GAE \cite{kipf2016variational}: a non-probabilistic variant of the VGAE model.
    \item GraphSAGE \cite{hamilton2017inductive}: a GNN model on static graphs, which supports inductive representation learning on large graphs. 
    % The reason why we didn't choose spectral domain based GNNs(\eg, \cite{kipf2016semi, velivckovic2017graph, xu2018powerful}) is that they are not practical when dealing with large graphs, \eg, our Taobao datasets which has almost two million nodes.
\end{itemize}

(2) \emph{Temporal approaches}, which train models or node embedding vectors on the temporal graph formed before the testing time. 
\begin{itemize}[leftmargin=*]
    \item CTDNE \cite{nguyen2018continuous}: based on random walks, it is a network embedding method which learns time-respecting embedding from continuous-time dynamic networks.
    \item EvolveGCN \cite{pareja2020evolvegcn}: using RNN to evolve GCN parameters to capture the dynamic information of sequences of static graph snapshots.
    \item GraphSAGE+T: our implementation based on GraphSAGE. Specifically, when aggregating neighbors' information, it will consider the time decay effect, \ie, earlier neighbors will get smaller weights while more recent neighbors will get larger weights during aggregation.
    \item TGAT \cite{xu2020inductive}: it uses the self-attention mechanism to aggregate temporal-topological neighborhood features. Besides, based on Bochner's theorem, it encodes the event time as part of the node embedding vector. 
\end{itemize}

(3) \emph{Hawkes process-based approaches}, which train node embedding vectors using the temporal graph formed before the testing time, based on Hawkes process. 
\begin{itemize}[leftmargin=*]
    \item HTNE \cite{zuo2018embedding}: a network embedding method which integrates the Hawkes process into network embedding so as to capture the influence of historical neighbors on the current neighbors.
    \item MMDNE \cite{lu2019temporal}: a network embedding method with micro- and macro-dynamics. Specifically, the micro-dynamics describe the link formation process, while the macro-dynamics refer to the evolution pattern of the network scale.
\end{itemize}

For DeepWalk, Node2vec and CTDNE, we set their random walk sampling parameters, such as number of walks, walk length and window size according to their recommended settings, respectively.  For all network embedding methods, the node embedding dimension is 128, which tends to perform well empirically. For all GNN-based methods, we set the number of layers, the  node embedding  dimensions  and the learning rates to the same with our model TREND. 
For HTNE, MMDNE and EvolveGCN, we set their historical window size to 5 (\ie, only use the 5 most recent neighbors), given the empirical performance and efficiency considerations.
For efficiency reasons, we perform random neighborhood sampling on GraphSAGE and GraphSAGE+T, setting the sample size to 10 on CollegeMsg and cit-HepTh, 20 on Wikipedia, and 5 on Taobao (which is the most sparse graph), the same with TREND; we use the same sample size on TGAT, but sample the more recent neighbors with higher probability following its original design.
For EvolveGCN, we use the EvolveGCN-O version. For TGAT, we set the number of attention heads to 3. 
Other hyperparameters are chosen  empirically, following guidance from literature.

\section{Scalability Study}\label{app:scala}

\begin{figure}[ht]
   \centering
   \includegraphics[width=0.5\linewidth]{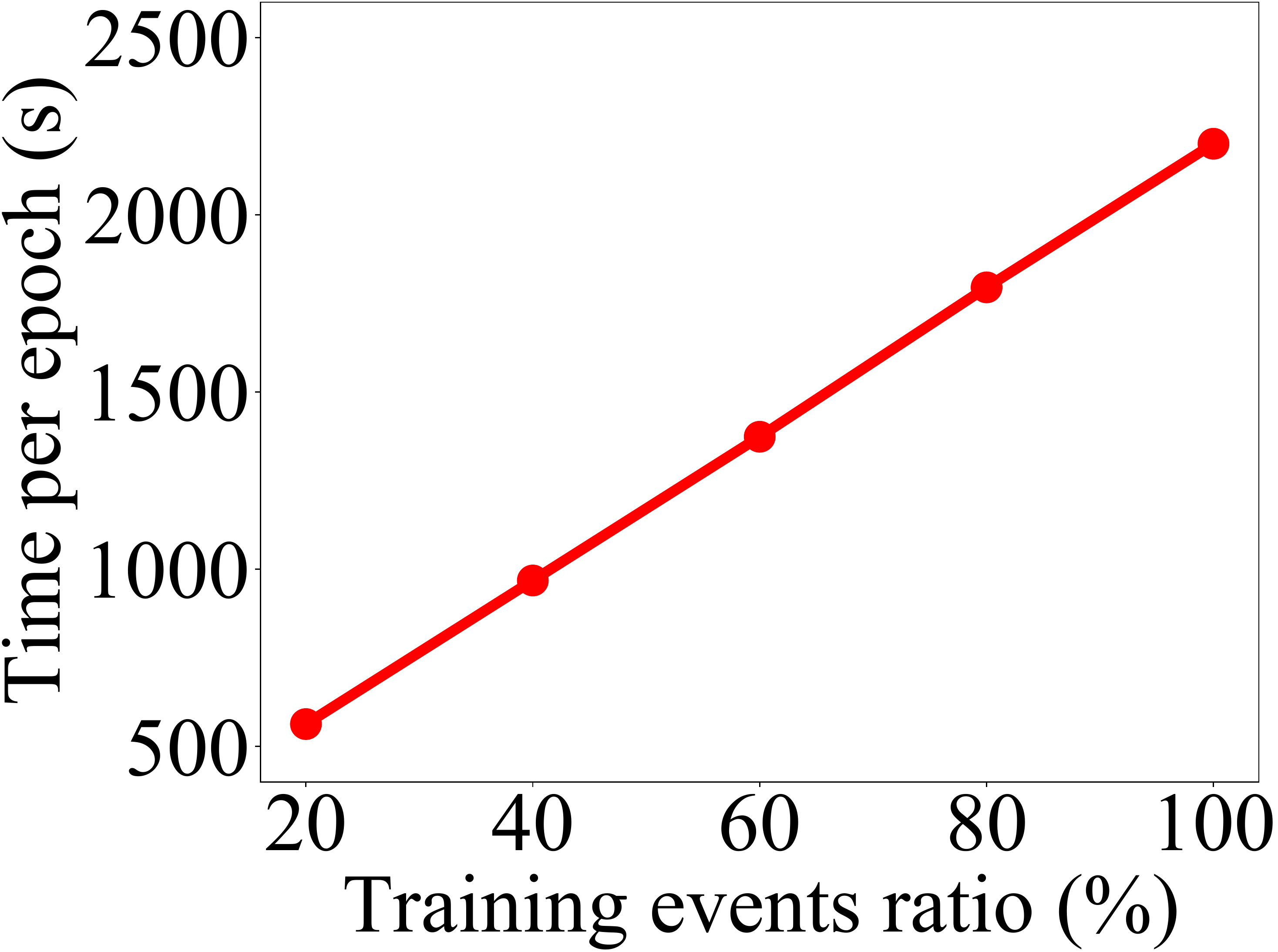}
   \caption{Time complexity.}
   \label{fig:time}
\end{figure}

On Taobao (with a total of more than 4 million events), we extract different ratios (20\%--100\%) of training events to form 5 subgraphs, and record the training time per epoch on each subgraph. In Fig.~\ref{fig:time}, the training time grows linearly in the number of training events, which is consistent with our complexity analysis, and implies that the proposed model is scalable.

% On Taobao (with more than 4 million events), we extract different ratio (from 20\% to 100\%) of training events to form 5 groups, and record time cost of different groups after running one epoch. As shown in Fig.~\ref{fig:time}, training time cost is linearly related to the number of training events, i.e., the proposed model is scalable.

%%
%% If your work has an appendix, this is the place to put it.
% \appendix

% \section{Research Methods}

% \subsection{Part One}

% Lorem ipsum dolor sit amet, consectetur adipiscing elit. Morbi
% malesuada, quam in pulvinar varius, metus nunc fermentum urna, id
% sollicitudin purus odio sit amet enim. Aliquam ullamcorper eu ipsum
% vel mollis. Curabitur quis dictum nisl. Phasellus vel semper risus, et
% lacinia dolor. Integer ultricies commodo sem nec semper.

% \subsection{Part Two}

% Etiam commodo feugiat nisl pulvinar pellentesque. Etiam auctor sodales
% ligula, non varius nibh pulvinar semper. Suspendisse nec lectus non
% ipsum convallis congue hendrerit vitae sapien. Donec at laoreet
% eros. Vivamus non purus placerat, scelerisque diam eu, cursus
% ante. Etiam aliquam tortor auctor efficitur mattis.

% \section{Online Resources}

% Nam id fermentum dui. Suspendisse sagittis tortor a nulla mollis, in
% pulvinar ex pretium. Sed interdum orci quis metus euismod, et sagittis
% enim maximus. Vestibulum gravida massa ut felis suscipit
% congue. Quisque mattis elit a risus ultrices commodo venenatis eget
% dui. Etiam sagittis eleifend elementum.

% Nam interdum magna at lectus dignissim, ac dignissim lorem
% rhoncus. Maecenas eu arcu ac neque placerat aliquam. Nunc pulvinar
% massa et mattis lacinia.

\end{document}